\documentclass[11pt,a4paper]{article}

\usepackage[T1]{fontenc}
\usepackage[utf8]{inputenc}
\usepackage{amsmath,amssymb}
\usepackage[dvipdfmx]{graphicx}
\usepackage{multirow}
\usepackage{bm}
\usepackage{float}
\usepackage{braket}
\usepackage{cite}
\usepackage{caption}
\usepackage{url}
\usepackage{hyperref}
\usepackage{here}
\usepackage{color}
\usepackage[top=3cm, bottom=3cm, left=2.5cm, right=2.5cm]{geometry}
\hypersetup{
	colorlinks=true,
	linkcolor=cyan,
	citecolor=blue
}

\makeatletter

\@addtoreset{equation}{section}
\makeatother

\allowdisplaybreaks[1]

\makeatother

\title{
	{\Large \bf  Correspondence  of topological classification 
		\\
		between quantum graph extra dimension and topological matter}
	\\*[20pt]
}

\author{
	Tomonori Inoue$^{a,}$\footnote{E-mail: \url{t-inoue@stu.kobe-u.ac.jp}}\ ,\ \ 
	Makoto Sakamoto$^{a,}$\footnote{E-mail: \url{dragon@kobe-u.ac.jp}}\ ,\ \ 
	Masatoshi Sato$^{b,}$\footnote{E-mail: \url{msato@yukawa.kyoto-u.ac.jp}}\ ,\ \  and\ \  
	Inori Ueba$^{c,}$\footnote{E-mail: \url{ueba@tomakomai.kosen-ac.jp}}\\*[20pt]
	 ${}^{(a)}${\it Department of Physics, Kobe University,  Kobe 657-8501, Japan}\\
	${}^{(b)}${\it Center for Gravitational Physics and Quantum Information,}\\{\it Yukawa Institute for Theoretical Physics, Kyoto University,}\\{\it
			Kyoto 606-8502, Japan}
		\\
		${}^{(c)}${\it National Institute of Technology, Tomakomai College,}\\{\it
			443 Nishikioka, Tomakomai 059-1275, Japan}
}

\date{}

\begin{document}
\begin{titlepage}
\centering
\maketitle
\begin{flushright}
\vspace{-10cm}
 {KOBE-TH-22-03}\\
 {YITP-22-33}\\
 \vspace{8.5cm}
\end{flushright}
\begin{abstract}
\normalsize
 In this paper, we study five-dimensional Dirac fermions of which extra-dimension is compactified on quantum graphs.
 We find that there is a non-trivial correspondence between matrices specifying boundary conditions at the vertex of the quantum graphs and zero-dimensional Hamiltonians in gapped free-fermion systems. 
 Based on the correspondence, we
 provide a complete topological classification of the boundary conditions
 in terms of non-interacting fermionic topological phases.
 The ten symmetry classes of topological phases are fully identified in the language of five-dimensional Dirac fermions, and topological numbers of the boundary conditions are given.
 In analogy with the bulk-boundary correspondence in non-interacting fermionic topological phases,
 the boundary condition topological numbers predict four-dimensional massless fermions localized at the vertex of the quantum graphs and thus govern the low energy physics in four dimensions.
\end{abstract}
\thispagestyle{empty}
\end{titlepage}
\newpage
\tableofcontents
\newpage
\section{Introduction}
\label{sec:introduction}
\begin{figure}[h]
	\centering
	\begin{minipage}{0.4\hsize}
	\centering
	\includegraphics[height=4.5cm]{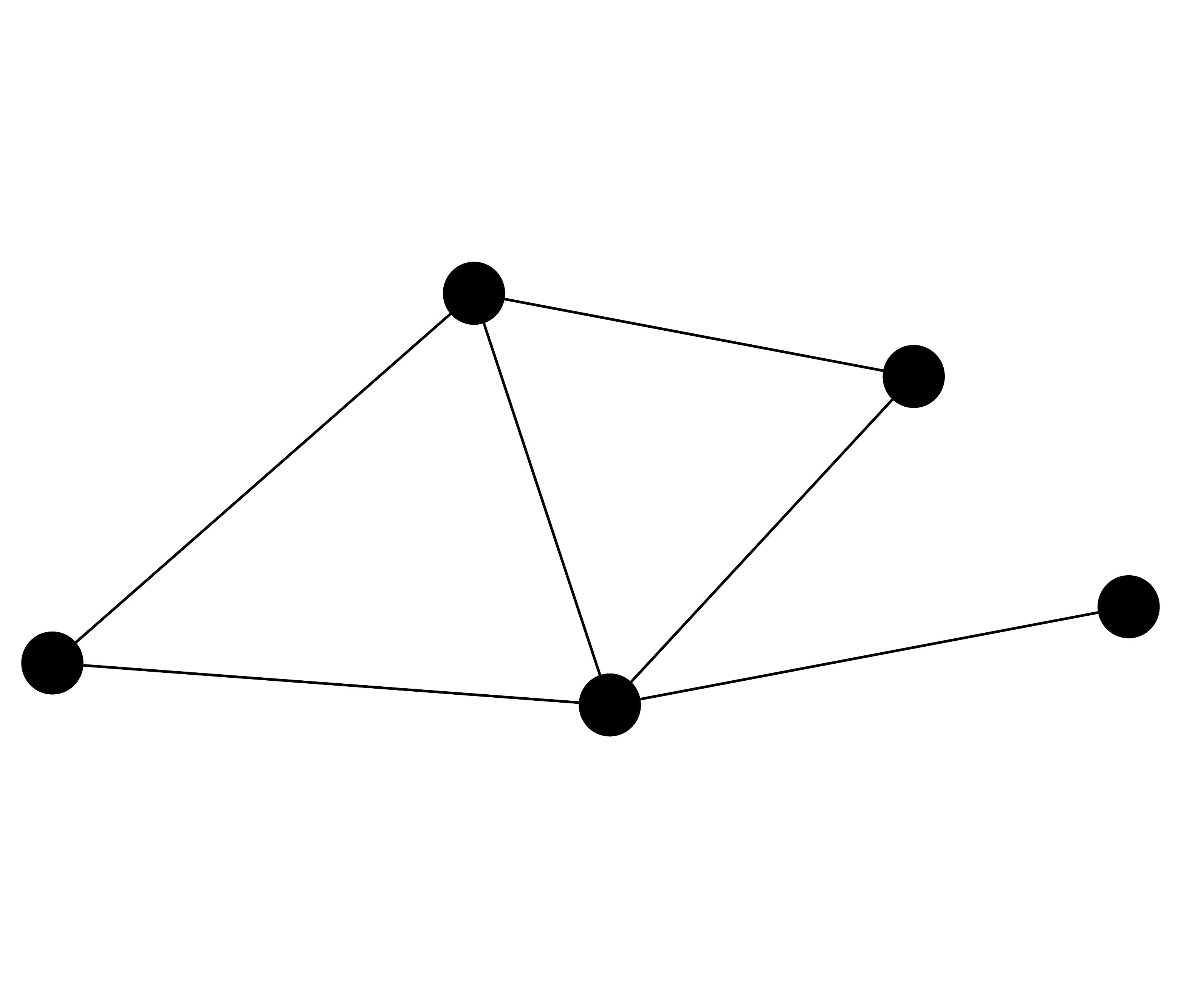}
	\caption{Quantum graph consisting of five vertices and six edges.}
	\label{fig:quantum-graph}
	\end{minipage}
\hspace{2cm}
\begin{minipage}{0.4\hsize}
	\centering
	\includegraphics[height=4.5cm]{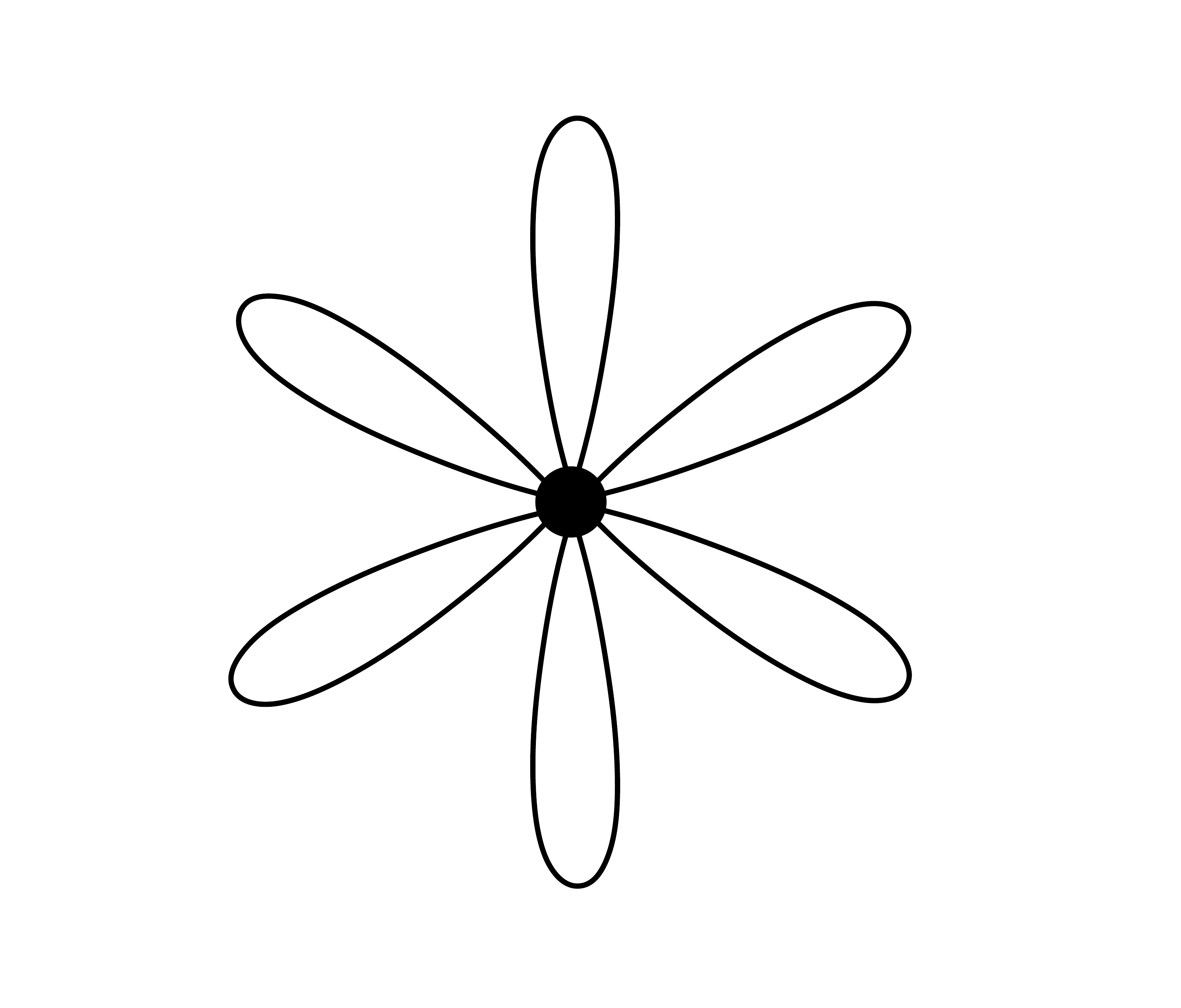}
	\caption{Rose graph consisting of one vertex and six loops.}
	\label{fig:rose-graph_1}
\end{minipage}
\end{figure}

A quantum graph is a one-dimensional (1d) graph that consists of edges and vertices connected to each other with differential operators defined on each edge like Figure~\ref{fig:quantum-graph} (see~\cite{Kuchment_2004,Kuchment_2005} for a review of the quantum graph). The quantum graph describes quantum mechanics on a 1d graph and has been paid attention to since we can obtain attractive physics caused by the degrees of freedom in boundary conditions for wave functions imposed at vertices from the requirement of current conservation.
The graph has been applied to the wide range of research areas, e.g. scattering theory on 1d graphs~\cite{Kostrykin:1998gz,Texier_2001,Boman:2005,Fujimoto:2018lzq}, quantum chaos~\cite{Kottos:1997,Cheon:2006,Gnutzmann:2010}, anyons~\cite{Harrison_2014,Maciazek:2017jon,Maci_ek_2019}, supersymmetric quantum mechanics~\cite{Nagasawa:2003tw,Nagasawa:2005kv,Ohya:2012qz}, Berry's phases~\cite{Ohya:2014ska,Ohya:2015xya,Ohya:2020lzi,Inoue:2021euj}, extra-dimensional models~\cite{Kim:2005aa,Cacciapaglia:2006tg,Bechinger:2009qk,Abel:2010kw,Law:2010pv,Fujimoto:2012wv,Fujimoto:2013ki,Fujimoto:2014fka,Fujimoto:2017lln,Fujimoto:2019lbo,Nortier:2020lbs}  and so on. 

Here we focus on applying quantum graphs to the extra space of 5d fermions, that is, 5d fermions with the extra space given by the quantum graph.
In the previous paper\cite{Fujimoto:2019fzb}, three of the present authors and collaborators revealed that 5d fermions on quantum graphs could naturally solve the problems of the fermion generation, namely the fermion mass hierarchy and the origin of the CP-violating phase in the standard model. 
However, the possible 4d effective theories remained unclear because the parameter space of the boundary conditions was huge.
Thus, the next step is a systematic investigation of the boundary conditions. 
A hint to this step is that we obtained 4d chiral zero modes localized at the vertex depending on the topological structure of the parameter space of the boundary conditions. This reminds us of topological insulators and superconductors, where gapless states appear on the boundaries by the topology in bulk from the bulk-boundary correspondence.
These topological matters are  classified into ten symmetry classes by the presence or absence of time-reversal, particle-hole, and chiral symmetries~\cite{Schnyder:2008tya,Kitaev:2009mg,Ryu:2010zza}.

In this paper, we perform a topological classification of the boundary conditions for 5d Dirac fermions on quantum graphs.
The boundary conditions are classified into ten symmetry classes by considering time reversal, charge conjugation, and extra spatial symmetries in 5d Dirac fermions. 
Surprisingly,  the classification of the boundary conditions  has a complete correspondence to the topological classification of 0+1d gapped free-fermion systems:
A Hermitian matrix that specifies the boundary conditions of quantum graphs  corresponds to a zero-dimensional Hamiltonian for topological insulators and superconductors.
In addition, symmetry classes of the boundary conditions coincide with those in the topological matter side. This means that the restrictions for the boundary conditions by the symmetries of 5d fermions are the same as those for zero-dimensional Hamiltonians with the Altland-Zirnbauder symmetries.
Furthermore,  we obtain the topological numbers $\mathbb{Z},\mathbb{Z}_2,2\mathbb{Z}$ for each symmetry class of the boundary conditions in the same manner as those of topological insulators and superconductors. 
Importantly, these numbers
predict the number of 4d massless fields localized at the vertex of quantum graphs, {\it i.e.}
the $\mathbb{Z}$ and $2\mathbb{Z}$ indices are equal to the number of Kaluza-Klein (KK) chiral zero modes,
and the $\mathbb{Z}_2$ index coincides with the parity of the number of Dirac zero modes.  
These zero modes would be regarded as gapless boundary states of topological phases.
A similar relation between boundary conditions and $1+1$d symmetry-protected topological (SPT) phases~\cite{Cho:2016xjw} or $1+2$d SPT phases~\cite{Han:2017hdv} has been known in the boundary conformal field theories.
In this paper, 
we reveal for the first time the relation between the boundary conditions on quantum graphs and $1+0$d SPT phases.

\begin{figure}[h]
		\centering
		\includegraphics[width=14cm]{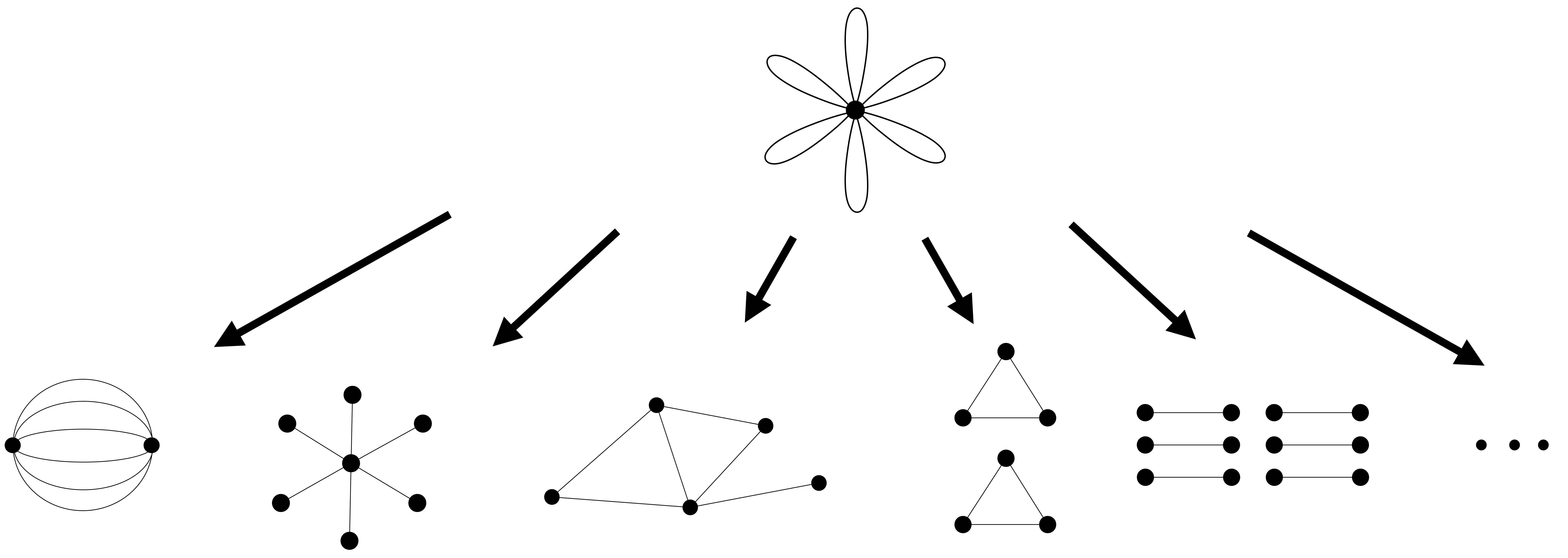}
		\caption{Decompositions of the rose graph with six edges.}
		\label{fig:rose-graph_decomosition}
\end{figure}

This paper is  organized as follows: In the next section, we briefly review 5d Dirac fermions on quantum graphs. 
Quantum graphs provide the most general 1d extra compactified spaces. 
In particular, we consider the so-called rose graph where any edge of the graph forms a loop that begins and ends at a common vertex, as shown in  Figure~\ref{fig:rose-graph_1}.
This choice does not lose the generality because
the rose graph can generate an arbitrary graph with the same number of edges by imposing suitable boundary conditions for wavefunctions at the vertex. See Figure~\ref{fig:rose-graph_decomosition}.\footnote{
	For example, if we impose the Dirichlet boundary condition on all edges of a rose graph,  wave functions on different edges are independent of each other. Thus, this boundary condition decomposes the rose graph into isolated edges  with the Dirichlet boundary conditions at both ends.
	The other graphs can also be obtained by suitable boundary conditions.
	Thus, we can collectively investigate any 1d extra space with $N$ edges by using a single rose graph with $N$ edges.} 
In Section \ref{sec:Tenfold classification of boundary conditions with symmetries}, we classify allowed boundary conditions in the rose graph subject to symmetries of 5d fermions and obtain ten symmetry classes. Then we relate them to Hamiltonians and symmetries in the gapped free-fermion systems.
In Section \ref{sec:Index in each symmetry classes}, we investigate the topological number of the boundary conditions in each symmetry class and the number of KK zero modes. 
Section \ref{sec:Summary and Discussion} is devoted to summary and discussion.
\vspace{-0.3cm}

\section{5d Dirac fermion on quantum graph}

\begin{figure}[h]
   	\centering
   	\includegraphics[height=6cm]{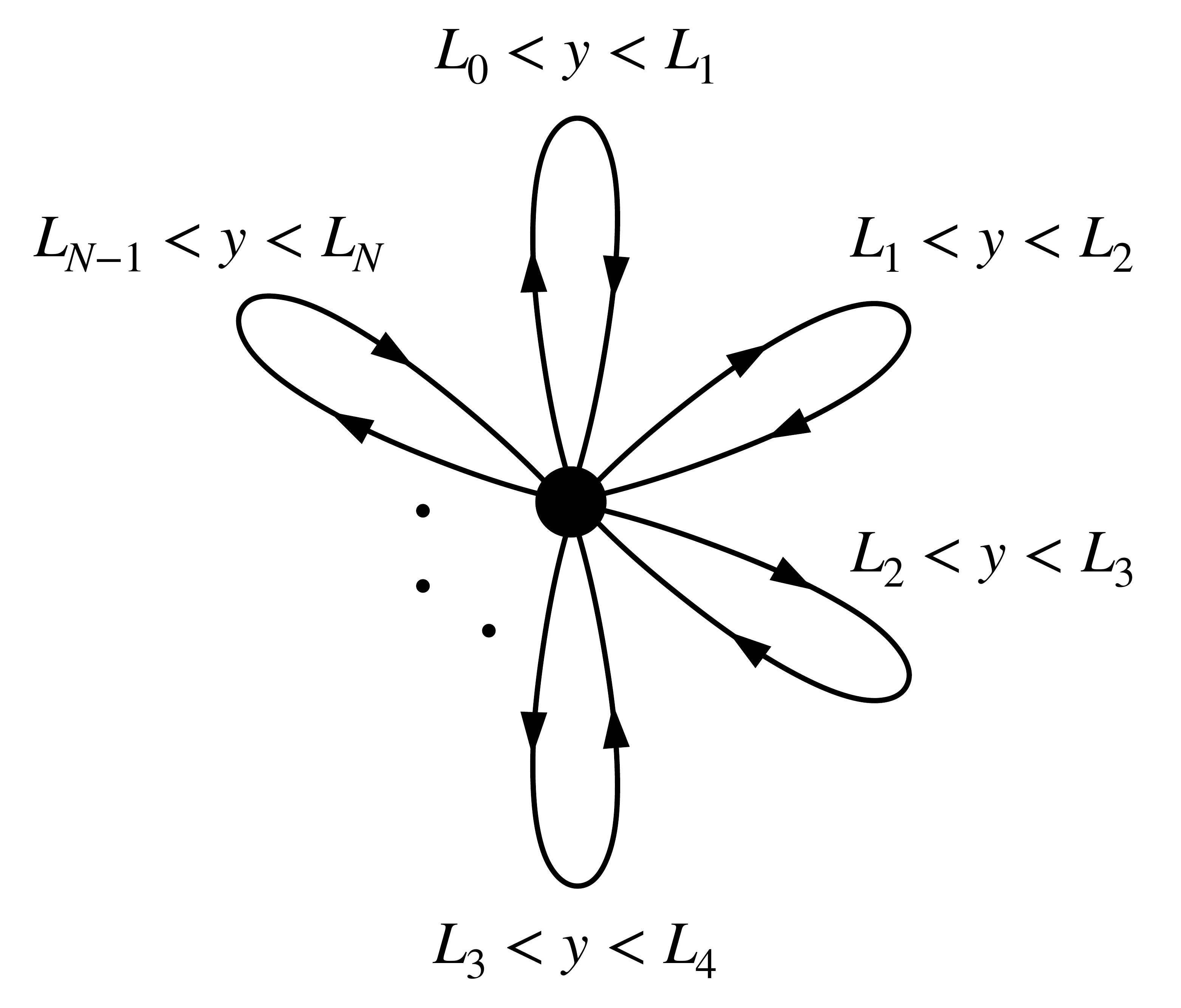}
   	\caption{Rose graph consisting of one vertex and $N$ loops.}
   	\label{fig:rose-graph}
   \end{figure}

In this section, we briefly review the properties of a 5d Dirac fermion on a quantum graph given in the previous paper\cite{Fujimoto:2019fzb}.
As discussed in section \ref{sec:introduction}, we take the extra space as the rose graph which consists of one vertex and $N$ loops with the length $L_a\ (a=1,\cdots,N)$  shown in Figure \ref{fig:rose-graph}. 
We  consider a KK decomposition of the 5d field and derive boundary conditions
that the field should satisfy at the vertex in the  graph. We also discuss how the degeneracy of
4d chiral zero-modes depend on the boundary conditions and show that it corresponds to the topological invariant  so-called Witten index.

\subsection{Setup}

Let us consider the 5d Dirac action 
	\begin{align}
	    S=\int d^{4}x\sum_{a=1}^{N}\int_{L_{a-1}}^{L_{a}}dy\:\overline{\Psi}(x,y)\left[i\gamma^{\mu}\partial_{\mu}+i\gamma^{y}\partial_{y}+M\right]\Psi(x,y)\,,
	    \label{eq:2-5daction}
	\end{align}
where $x^{\mu}$s' $(\mu=0,1,2,3)$ are the coordinates of the 4d Minkowski space-time and $y$ is the coordinate on the rose graph. $\Psi(x,y)$ is a four-component 5d Dirac spinor and the Dirac conjugate $\overline{\Psi}$ is defined by $\overline{\Psi}=\Psi^{\dagger}\gamma^{0}$.
$\gamma^{\mu}$s' $(\mu=0,1,2,3)$ are $4\times4$ gamma matrices that satisfy the Clifford algebra
	\begin{align}
	   \{\gamma^{\mu},\gamma^{\nu}\}=-2\eta^{\mu\nu},\quad\eta^{\mu\nu}=\mathrm{diag}(-,+,+,+)\,,
	    \label{eq:2-Clifford}
	\end{align}
and $\gamma^{y}$ is taken to be $\gamma^{y}=-i\gamma^{5}\:(\gamma^{5}=i\gamma^{0}\gamma^{1}\gamma^{2}\gamma^{3})$ with 4d chiral matrix $\gamma^5$.
The hermiticity of the gamma matrices is given by
	\begin{align}
	   (\gamma^{0})^{\dagger}=\gamma^{0}\,,\quad(\gamma^{i})^{\dagger}=-\gamma^{i}\ \ (i=1,2,3)\,,\quad(\gamma^{y})^{\dagger}=-\gamma^{y}\,.
	\end{align}
The parameter $M$ in the action \eqref{eq:2-5daction} is the bulk mass of the 5d Dirac fermion. 


It should be noted that the model in Eq.(\ref{eq:2-5daction}) can realize the domain wall fermion used in the lattice gauge theory~\cite{Kaplan:1992bt} as a special case.
The domain wall configuration is obtained by a circular rose graph consisting of two edges and two vertices, where the bulk mass has an opposite sign on the two edges.
Each vertex of this graph supports a 4d chiral fermion, of which chirality is opposite to each other on the two vertices. 
In the long length limit of the edges, the two 4d chiral fermions are decoupled to each other, so a chiral theory is realized effectively. 
Because rose graphs allow configurations other than the domain wall, 
our quantum graph approach is  more general than that of the domain wall fermions.

The action principle $\delta S=0$ gives the 5d Dirac equation
	\begin{align}
			\left[i\gamma^\mu\partial_\mu+i\gamma^y\partial_y+M\right]\Psi(x,y)=0\,,
	\end{align}
and also the condition for the surface term
	\begin{align}
		\sum_{a=1}^{N}\left[\, \overline{\Psi}(x,y)\gamma^{y}\delta\Psi(x,y)\right]_{y=L_{a-1}+\varepsilon}^{y=L_{a}-\varepsilon}=0\,,
		\label{eq:2-ConservationLaw}
	\end{align}
where $\varepsilon$ is an infinitesimal positive constant. This condition can be regarded as the conservation of  the current for $y$-direction. As we will see in the next section, Eq.\:(\ref{eq:2-ConservationLaw}) leads to boundary conditions that the field $\Psi(x,y)$ should obey at
the boundaries $y=L_{0}+\varepsilon,L_{1}\pm\varepsilon,\cdots,L_{N-1}\pm\varepsilon,L_{N}-\varepsilon$.\footnote
{
	If we impose boundary conditions such that partial summations of the terms in the left hand side of Eq.\:\eqref{eq:2-ConservationLaw} vanish independently, the rose graph is decomposed into graphs correspond to the partial summations, as discussed in Section \ref{sec:introduction}.
}

If the extra dimension is compact, a higher dimensional field can be decomposed into 4d fields with $x$-dependence and  Kaluza--Klein (KK) mode functions  with discrete eigenvalues on an extra space. Here, we decompose $\Psi(x,y)$ into 4d chiral fields $\psi_{\mathrm{R/L},n}^{(i)}(x)$ and KK mode functions $f_{n}^{(i)}(y)\,,\ g_{n}^{(i)}(y)$ as follows:
	\begin{align}
	    \Psi(x,y)=\sum_{i}\sum_{n}\left[\psi_{\mathrm{R},n}^{(i)}(x)f_{n}^{(i)}(y)+\psi_{\mathrm{L},n}^{(i)}(x)g_{n}^{(i)}(y)\right]\,,
	    \label{eq:2-KK_expantion1}
	\end{align}
where the index $n$ indicates the $n$-th level of the KK modes and $i$ denotes the index that distinguishes the degeneracy of the $n$-th KK modes (if it exists). The mode functions $f_{n}^{(i)}(y)$ and $g_{n}^{(i)}(y)$ are assumed to form a complete set, respectively, and satisfy the orthonormality relations
	\begin{align}
	    \sum_{a=1}^{N}\int_{L_{a-1}}^{L_{a}}dy\:f_{n}^{(i)*}(y)f_{m}^{(j)}(y)
	    & =\delta_{nm}\delta^{ij}\,,
	    \label{eq:2-ModeRelations1}\\
	    \sum_{a=1}^{N}\int_{L_{a-1}}^{L_{a}}dy\:g_{n}^{(i)*}(y)g_{m}^{(j)}(y)
	    & =\delta_{nm}\delta^{ij}\,.
	    \label{eq:2-ModeRelations2}
	\end{align}

We also require that the 4d fields $\psi_{\mathrm{R/L},n}^{(i)}$ are mass eigenstates with masses $m_n$ and  the following  4d action can be obtained by substituting the decomposition (\ref{eq:2-KK_expantion1}) into the 5d action Eq.\:(\ref{eq:2-5daction}) :
	\begin{align}
		S & =
		\int d^{4}x\Bigl\{\sum_{i}\overline{\psi}_{\mathrm{R},0}^{(i)}(x)i\gamma^{\mu}\partial_{\mu}\psi_{\mathrm{R},0}^{(i)}(x)+\sum_{j}\overline{\psi}_{\mathrm{L},0}^{(j)}(x)i\gamma^{\mu}\partial_{\mu}\psi_{\mathrm{L},0}^{(j)}(x)\nonumber \\
		& \qquad\qquad\qquad+\sum_{i}\sum_{n\neq0}\overline{\psi}_{n}^{(i)}(x)\left(i\gamma^{\mu}\partial_{\mu}+m_{n}\right)\psi_{n}^{(i)}(x)\Bigr\}\,,
	\end{align}
where $\psi_{n}^{(i)}(x)\equiv\psi_{\mathrm{R},n}^{(i)}(x)+\psi_{\mathrm{L},n}^{(i)}(x)$
for $n\neq0$, and $\psi_{\mathrm{R/L},0}^{(i)}(x)$ denote the 4d
chiral massless spinors with $m_{0}=0$.
Then the mode functions $f_{n}^{(i)}(y)$ and $g_{n}^{(i)}(y)$ should satisfy the relation
	\begin{align}
	    (\partial_{y}+M)f_{n}^{(i)}(y)
	     & =m_{n}g_{n}^{(i)}(y)\,,
	     \label{eq:2-SUSYrelations1}\\
	    (-\partial_{y}+M)g_{n}^{(i)}(y)
	     & =m_{n}f_{n}^{(i)}(y)\,.
	     \label{eq:2-SUSYrelations2}
	\end{align}
It follows that $f_{n}^{(i)}$ and $g_{n}^{(i)}$ are given by the eigenfunctions of the
equations
	\begin{align}
	    (-\partial_{y}^{2}+M^{2})f_{n}^{(i)}(y)
	     & =m_{n}^{2}f_{n}^{(i)}(y)\,,
	     \label{eq:2-eigenequation_1}\\
	    (-\partial_{y}^{2}+M^{2})g_{n}^{(i)}(y)
	     & =m_{n}^{2}g_{n}^{(i)}(y)\,,
	     \label{eq:2-eigenequation_2}
	\end{align}
and the 4D masses $m_n$ are determined by solving these equations with allowed boundary conditions.

\subsection{Boundary conditions}
\label{subsec:Boundary conditions}

Then, let us derive the allowed boundary conditions. From the decomposition
(\ref{eq:2-KK_expantion1}) of $\Psi(x,y)$ (and also $\delta\Psi(x,y)$) and the independence of the fields $\psi_{\mathrm{R/L},n}^{(i)}(x)$ and $\delta\psi_{\mathrm{R/L},n}^{(i)}(x)$, Eq.\:(\ref{eq:2-ConservationLaw}) can be written as
  \begin{align}
    \vec{F}_{n}^{(i)\dagger}\vec{G}_{m}^{(j)}=\vec{G}_{m}^{(j)\dagger}\vec{F}_{n}^{(i)}=0\quad\text{for}\quad\forall n,m,i,j,
    \label{eq:3-orthogonal-condition}
  \end{align}
where $\vec{F}_{n}^{(i)}$ and $\vec{G}_{m}^{(j)}$ are $2N$-dimensional
complex vectors defined by
  \begin{align}
    \vec{F}_{n}^{(i)}\equiv
      \begin{pmatrix}f_{n}^{(i)}(L_{0}+\varepsilon)\\
      f_{n}^{(i)}(L_{1}-\varepsilon)\\
      f_{n}^{(i)}(L_{1}+\varepsilon)\\
      f_{n}^{(i)}(L_{2}-\varepsilon)\\
      \vdots\\
      f_{n}^{(i)}(L_{a-1}+\varepsilon)\\
      f_{n}^{(i)}(L_{a}-\varepsilon)\\
      \vdots\\
      f_{n}^{(i)}(L_{N-1}+\varepsilon)\\
      f_{n}^{(i)}(L_{N}-\varepsilon)
      \end{pmatrix}\,,
    \qquad\vec{G}_{m}^{(j)}\equiv
      \begin{pmatrix}g_{m}^{(j)}(L_{0}+\varepsilon)\\
      -g_{m}^{(j)}(L_{1}-\varepsilon)\\
      g_{m}^{(j)}(L_{1}+\varepsilon)\\
      -g_{m}^{(j)}(L_{2}-\varepsilon)\\
      \vdots\\
      g_{m}^{(j)}(L_{a-1}+\varepsilon)\\
      -g_{m}^{(j)}(L_{a}-\varepsilon)\\
      \vdots\\
      g_{m}^{(j)}(L_{N-1}+\varepsilon)\\
      -g_{m}^{(j)}(L_{N}-\varepsilon)
      \end{pmatrix}\,.
    \label{eq:boundary-vectors}
  \end{align}
These vectors consist of values of the mode function at the boundaries $y=L_{0}+\varepsilon,L_{1}\pm\varepsilon,\cdots,L_{N-1}\pm\varepsilon,L_{N}-\varepsilon$.
Here, we call $\vec{F}_{n}^{(i)}$and $\vec{G}_{m}^{(j)}$ boundary vectors.
The condition \eqref{eq:3-orthogonal-condition} means that the vector space spanned by $\{\vec{F}_{n}^{(i)}\}$ is orthogonal to the one by $\{\vec{G}_{m}^{(j)}\}$\,.

Now, to solve the equations \eqref{eq:2-eigenequation_1} and \eqref{eq:2-eigenequation_2} and determine the mass eigenvalues $m_n$,  we want to obtain $4N$ constraints in total, since the graph has $2N$ boundaries $y=L_{0}+\varepsilon,L_{1}\pm\varepsilon,\cdots,L_{N-1}\pm\varepsilon,L_{N}-\varepsilon$ for $f_{n}^{(i)}$ and $g_{m}^{(j)}$.
However, the relations \eqref{eq:2-SUSYrelations1} and \eqref{eq:2-SUSYrelations2} imply that the massive mode functions $f_{n}^{(i)}$ and $g_{m}^{(j)}$ are related with each other (except for the zero mode functions which obey the first differential equations). Thus, we should require that the condition \eqref{eq:3-orthogonal-condition} provides $2N$ constraints in total, otherwise the system is undetermined or overdetermined.\footnote{For example, if we impose $4N$ constraints with the condition that $f_{n}^{(i)}$ and $g_{m}^{(j)}$ equal to 0 at each boundary, there are no solutions for \eqref{eq:2-eigenequation_1} and \eqref{eq:2-eigenequation_2} since the mode functions should satisfy both the Dirichlet and Neumann boundary conditions from the relations \eqref{eq:2-SUSYrelations1} and \eqref{eq:2-SUSYrelations2}.}
It follows from this observation that Eq.\:(\ref{eq:3-orthogonal-condition}) is replaced by the boundary conditions
	\begin{align}
	    (1_{2N}-U_{\mathrm{B}})\vec{F}_{n}^{(i)}
	     & =0\,,
	    \label{eq:BC-F}\\
	    (1_{2N}+U_{\mathrm{B}})\vec{G}_{m}^{(j)}
	     & =0\,,
	    \label{eq:BC-G}
	\end{align}
where $U_{\mathrm{B}}$ is a $2N\times2N$ Hermitian unitary matrix
	\begin{align}
		U_{\mathrm{B}}^{2}=1_{2N}\,,\quad 		U_{\mathrm{B}}^{\dagger}=U_{\mathrm{B}}\,.\label{eq:U2-condition}
	\end{align}
We can find that $(1_{2N}\mp U_{\mathrm{B}})/2$  in Eqs.\:\eqref{eq:BC-F} and \eqref{eq:BC-G} correspond to the projection matrices which map the $2N$-dimensional complex vector space into the ones spanned by $\{\vec{F}_{n}^{(i)}\}$ and $\{\vec{G}_{m}^{(j)}\}$, respectively.
 Thus, we conclude that a Hermitian unitary matrix $U_{\mathrm{B}}$ specifies a 5d Dirac theory on a rose graph depicted in Figure \ref{fig:rose-graph}. 

We can classify the matrix $U_{\mathrm{B}}$ into $2N+1$ types by the number of the eigenvalues $+1$ (or $-1$).
We call the case with $k$ negative eigenvalues  the type ($2N-k,k$)
boundary condition (BC) ($k=0,1,\cdots,2N$).  The matrix $U_{\mathrm{B}}$ in this type can
be represented as
  \begin{align}
    \text{Type}(2N-k,k): && U_{\mathrm{B}}=V
    \left(\begin{array}{ccc|ccc}
    +1 &  & {0}\\
     & \ddots &  &  & {0}\\
    {0} &  & +1\\
    \hline  &  &  & -1 &  & {0}\\
     & {0} &  &  & \ddots\\
     &  &  & {0} &  & -1
    \end{array}\right)
    V^{\dagger}&&(0\leq k\leq2N)
    \label{eq:type(2N-k,k)}\\
     && \  \underbrace{\hspace{2.4cm}}_{2N-k}\underbrace{\hspace{2.4cm}}_{k}\hspace{0.85cm}\nonumber&&
  \end{align}
with a $2N\times2N$ unitary matrix $V$. Therefore the parameter space of the type ($2N-k,k$) BC is given by the complex Grassmaniann
  \begin{align}
    \frac{U(2N)}{U(2N-k)\times U(k)}\,.
  \end{align}
Since continuous deformations of $V$ do not change the numbers
of positive and negative eigenvalues in $U_{\mathrm{B}}$, the different
types of boundary conditions cannot be connected continuously.

For later convenience, we write the $2N\times2N$ unitary matrix $V$
as
  \begin{align}
    V=(\vec{u}_{1},\:\vec{u}_{2},\:\cdots\:,\vec{u}_{2N})
    \label{eq:matrix-V}
  \end{align}
where $\vec{u}_{r}\:(r=1,2,\cdots,2N)$ are $2N$-dimensional orthonormal
complex vectors which satisfy $\vec{u}_{r}^{\dagger}\vec{u}_{s}=\delta_{rs}\:(r,s=1,2,\cdots,2N)$.
Then, the matrix $U_{\mathrm{B}}$ for the type ($2N-k,k$) BC can be expressed by
  \begin{align}
    U_{\mathrm{B}}=\sum_{r=1}^{2N-k}\vec{u}_{r}\vec{u}_{r}^{\dagger}-\sum_{r=2N-k+1}^{2N}\vec{u}_{r}\vec{u}_{r}^{\dagger}
  \end{align}
and the boundary conditions (\ref{eq:BC-F}) and (\ref{eq:BC-G})
are of the form
  \begin{align}
    \vec{u}_{r}^{\dagger}\vec{F}_{n}^{(i)}
     & =0\quad\text{for}\quad r=2N-k+1,\cdots,2N\,,
    \label{eq:BC'-F}\\
    \vec{u}_{r}^{\dagger}\vec{G}_{m}^{(j)}
     & =0\quad\text{for}\quad r=1,\cdots,2N-k\,.
    \label{eq:BC'-G}
  \end{align}

\subsection{Zero-mode degeneracy}
One of the features of the quantum graph is  that the  mode functions can be degenerate due to the boundary conditions. 
In our model, this degeneracy can be regarded as the number of four dimensional fields with degenerate masses.
In particular, the degeneracy of zero mode solutions plays important roles to classify the boundary conditions.
Then,  we focus on the zero mode solutions $f_{0}^{(i)}(y)$
and $g_{0}^{(j)}(y)$, which obey the equations (see Eqs.\: (\ref{eq:2-SUSYrelations1})
and (\ref{eq:2-SUSYrelations2}))
  \begin{align}
    (\partial_{y}+M)f_{0}^{(i)}(y)
     & =0\,,\\
    (-\partial_{y}+M)g_{0}^{(j)}(y)
     & =0\,.
  \end{align}
The mode functions $f_{0}^{(i)}(y)$ and $g_{0}^{(j)}(y)$ on the
rose graph can be discontinuous at the vertex and written into the form of exponentially localized functions:
  \begin{align}
    f_{0}^{(i)}(y)
     & =\sum_{a=1}^{N}\theta(y-L_{a-1})\theta(L_{a}-y)F_{a}^{(i)}C_{a}e^{-My}\,,\label{eq:zero mode f}\\
    g_{0}^{(j)}(y)
     & =\sum_{a=1}^{N}\theta(y-L_{a-1})\theta(L_{a}-y)G_{a}^{(i)}C_{a}'e^{+My}\,,
     \label{eq:zero mode g}
  \end{align}
where $\theta(y)$ denotes the Heaviside step function and the complex
constants $F_{a}^{(i)},\:G_{a}^{(j)}\in\mathbb{C}\:(a=1,\cdots,N)$
are determined by the boundary conditions. Here we also introduced
the constants 
  \begin{align}
    C_{a}=\sqrt{\frac{1}{e^{-2M(L_{a-1}-\varepsilon)}+e^{-2M(L_{a}+\varepsilon)}}}\,,\quad C_{a}'=\sqrt{\frac{1}{e^{2M(L_{a-1}-\varepsilon)}+e^{2M(L_{a}+\varepsilon)}}}\,,
    \label{eq:C,C'}
  \end{align}
for later convenience.

We can see that the number of linearly independent solutions $f_{0}^{(i)}(y)\:(g_{0}^{(j)}(y))$ corresponds to the number of linearly independent $N$-dimensional complex
vectors $\bm{F}^{(i)}\equiv(F_{1}^{(i)},F_{2}^{(i)},\cdots,F_{N}^{(i)})^{\top}$ 
$(\bm{G}^{(j)}\equiv(G_{1}^{(j)},G_{2}^{(j)},\cdots,G_{N}^{(j)})^{\top})$.
Then, let us discuss the vectors $\bm{F}^{(i)}$ and $\bm{G}^{(j)}$ under the type $(2N-k,k)$
BC.
For this purpose, we introduce orthonormal $2N$-dimensional
vectors $\vec{\mathcal{F}}_{a}$ and $\vec{\mathcal{G}}_{a}\:(a=1,\cdots,N)$ 
  \begin{align}
    \vec{\mathcal{F}}_{a} & \equiv C_{a}(\underbrace{0,\cdots,0}_{2(a-1)},e^{-M(L_{a-1}+\varepsilon)},e^{-M(L_{a}-\varepsilon)},0,\cdots,0)^{\top}\,,
    \label{eq:base-F}\\
    \vec{\mathcal{G}}_{a} & \equiv C'_{a}(\underbrace{0,\cdots,0}_{2(a-1)},e^{M(L_{a-1}+\varepsilon)},-e^{M(L_{a}-\varepsilon)},0,\cdots,0)^{\top}\,,
    \label{eq:base-G}
  \end{align}
which form a complete set in the $2N$-dimensional complex vector space.
Here the constants $C_{a}$ and $C'_{a}$ are the same as Eq.\:(\ref{eq:C,C'}).

Then the boundary vectors $\vec{F}_{0}^{(i)}$, $\vec{G}_{0}^{(j)}$
in Eq.\:(\ref{eq:boundary-vectors}) for $n=0$ and the $2N$-dimensional
complex vectors $u_{p}\:(p=1,\cdots,2N)$ in Eq.\:(\ref{eq:matrix-V}) can be decomposed by $\vec{\mathcal{F}}_{a}$ and $\vec{\mathcal{G}}_{a}\:(a=1,\cdots,N)$ as follows:
  \begin{align}
    \vec{F}_{0}^{(i)}
     & =\sum_{a=1}^{N}F_{a}^{(i)}\vec{\mathcal{F}}_{a}\,,\quad\vec{G}_{0}^{(j)}=\sum_{a=1}^{N}G_{a}^{(j)}\vec{\mathcal{G}}_{a}\,,\\
    \vec{u}_{r}
     & =\sum_{a=1}^{N}\alpha_{r,a}\vec{\mathcal{F}}_{a}+\sum_{a=1}^{N}\beta_{r,a}\vec{\mathcal{G}}_{a}\quad(r=1,2,\cdots,2N)\,,
     \label{eq:vector-u}
  \end{align}
where $\alpha_{r,a}$ and $\beta_{r,a}$ are complex constants and
satisfy
  \begin{align}
    \sum_{a=1}^{N}(\alpha_{r,a}^{*}\alpha_{s,a}+\beta_{r,a}^{*}\beta_{s,a})=\delta_{rs}\quad(r,s=1,2,\cdots,2N)
  \end{align}
from the orthonormal relations for $\vec{u}_{r}$.
From the boundary conditions (\ref{eq:BC'-F}) and (\ref{eq:BC'-G}) for
$n=0$, we obtain the conditions that $\bm{F}^{(i)}$ and $\bm{G}^{(j)}$ are orthogonal to  the vector
$\bm{\alpha}_{q}\equiv(\alpha_{q,1},\alpha_{q,2},\cdots,\alpha_{q,N})^{\top}$    $(q=2N-k+1,\cdots,2N)$ and
$\bm{\beta}_{p}\equiv(\beta_{p,1},\beta_{p,2},\cdots,\beta_{p,N})^{\top}\,(p=1,\cdots,2N-k)$, respectively:
  \begin{align}
    \bm{\alpha}_{q}^{\dagger}\bm{F}^{(i)}
     & =0\quad(q=2N-k+1,\cdots,2N)\,,
     \label{eq:BC''-F}\\
    \bm{\beta}_{p}^{\dagger}\bm{G}^{(j)}
     & =0\quad(p=1,2,\cdots,2N-k)\,.
     \label{eq:BC''-G}
  \end{align}

\begin{table}[t]
	\centering
	\captionsetup{width=.90\linewidth}
	\caption{The number of the zero mode solutions of $f_{0}^{(i)}(y)$ and $g_{0}^{(j)}(y)$
		for the type $(2N-k,k)$ BC. $\ell$ denotes the maximal
		number of the linearly independent vectors $\bm{\alpha}_{q}\:(q=2N-k+1,\cdots,2N)$
		in Eq.\:(\ref{eq:vector-u}). $N_{f_{0}}$ ($N_{g_{0}}$) is the number
		of the zero mode solutions of $f_{0}^{(i)}(y)$ ($g_{0}^{(j)}(y)$).
		We can find that $N_{f_{0}}-N_{g_{0}}$ is independent of $\ell$, though
		both of $N_{f_{0}}$ and $N_{g_{0}}$ depend of $\ell$.}
	\label{table:3-1}
	\begin{tabular}{c|cccc}
		\hline
		$k$ & $\ell$ & $N_{f_{0}}$ & $N_{g_{0}}$ & $\Delta_{\mathrm{W}}=N_{f_{0}}-N_{g_{0}}$\tabularnewline
		\hline
		\multirow{4}{*}{$0\leq k\leq N$} & 0 & $N$ & $k$ & $N-k$\tabularnewline
		& 1 & $N-1$ & $k-1$ & $N-k$\tabularnewline
		& $\vdots$ & $\vdots$ & $\vdots$ & $\vdots$\tabularnewline
		& $k$ & $N-k$ & 0 & $N-k$\tabularnewline
		\hline
		\multirow{4}{*}{$N\leq k\leq2N$} & $k-N$ & $2N-k$ & $N$ & $N-k$\tabularnewline
		& $k-N+1$ & $2N-k-1$ & $N-1$ & $N-k$\tabularnewline
		& $\vdots$ & $\vdots$ & $\vdots$ & $\vdots$\tabularnewline
		& $N$ & 0 & $k-N$ & $N-k$\tabularnewline
		\hline
	\end{tabular}
\end{table}

Let us suppose that the number of the linearly independent vectors for $\bm{\alpha}_{q}\:(q=2N-k+1,\cdots,2N)$ is $\ell$. 
Then the number of the linearly independent vectors for $\bm{\beta}_{q}\:(q=2N-k+1,\cdots,2N)$, $\bm{\alpha}_{p}$ and $\bm{\beta}_{p}\:(p=1,\cdots,2N-k)$ are $k-\ell,N-\ell$ and $N-k+\ell$, respectively, because of the independence of $\vec{u}_{p}\:(p=1,\cdots,2N-k)$ and $\vec{u}_{q}\:(q=2N-k+1,\cdots,2N)$.
Therefore the range of $\ell$
is restricted to $0\leq \ell\le k$ for the case of $k=0,\cdots,N$ and
$k-N\leq \ell\leq N$ for the case of $k=N,\cdots,2N$.
If the number of the linearly independent vectors for $\bm{\alpha}_{q}\:(q=2N-k+1,\cdots,2N)$
and $\bm{\beta}_{p}\:(p=1,\cdots,2N-k)$ are $\ell$ and $N-k+\ell$, respectively,
we can find that there exist $N-\ell$ linearly independent solutions
for $\bm{F}^{(i)}\:(i=1,\cdots,N-\ell)$ and $k-\ell$ linearly independent
solutions for $\bm{G}^{(j)}\:(j=1,\cdots,k-\ell$) from Eqs.\: (\ref{eq:BC''-F})
and (\ref{eq:BC''-G}). 
Therefore, for the type $(2N-k,k)$ BC, we can conclude
that the degeneracy of zero mode $f_{0}^{(i)}(y)$ is given by $N_{f_{0}}=N-\ell$,
and that of zero mode $g_{0}^{(i)}(y)$ is given by $N_{g_{0}}=k-\ell$.
The degeneracies $N_{f_{0}}$ and $N_{g_{0}}$ for each boundary condition
are described in Table \ref{table:3-1}. 

We also comment about phenomenological aspects of this model.
In the case of the type ($2N-k,k$) BC, there are $|N-k|$ massless chiral fields in the 4d effective theory since  pairs of left and right-handed chiral zero modes can form massless  Dirac spinors (and these may become massive through quantum corrections if we introduce interactions).  Therefore, we can obtain three generations of chiral fermions for the type ($N-3,N+3$) and ($N+3,N-3$) BCs. Since the zero mode functions are exponentially localized at some boundaries, overlap integrals of the mode functions can easily produce  hierarchical masses and also the flavor mixing if we introduce  the 5d Yukawa interactions. 
Moreover, complex parameters in the matrix $U_{\mathrm{B}}$ generally give the genuine complex zero mode functions and this would result in the CP violating phase in the CKM matrix.
Therefore, we can find that this model has the desired properties to explain the fermion flavor structure in the standard model from the viewpoint of higher dimensional theories.

\subsection{Witten index}
\label{subsec:Witten index}

We can notice that the difference $N_{f_{0}}-N_{g_{0}} (=N-k)$ is independent of $\ell$. This implies that the number of chiral zero modes is invariant under continuous deformations of the boundary conditions, since the type\,($2N-k,k$) BC is not continuously connected to the type\,($2N-k',k'$) one with $k\neq k'$ (although $\ell$ can be changed by those deformations). 
This topological property is related to a hidden structure of the supersymmetric quantum mechanics in this model.

If we introduce the two-component functions constructed from the mode functions
	\begin{align}
			\Phi_{n,+}^{(i)}(y)=\begin{pmatrix}
						f^{(i)}_{n}(y) \\ 0
				\end{pmatrix}\,,
			\quad
			\Phi_{n,-}^{(i)}(y)=\begin{pmatrix}
				0 \\g^{(i)}_{n}(y)
			\end{pmatrix}
	\end{align} 
and also the Hermitian operators $H$, $Q$ and $(-1)^{F}$ defined by
\begin{align}
	H=\begin{pmatrix}
		-\partial_{y}^2+M & 0\\ 0 & -\partial_{y}^2+M
		\end{pmatrix}\,,
		\quad 
		Q=\begin{pmatrix}
				0&-\partial_{y}+M\\ \partial_{y}+M&0
			\end{pmatrix}\,,
		\quad (-1)^{F}=\begin{pmatrix}1&0\\0&-1\end{pmatrix},
\end{align}
we can find that these satisfy the supersymmetric relations
	\begin{align}
		&H=Q^2\,,\quad  \{Q,(-1)^F\}=0\,,\quad [H,(-1)^F]=0\,, 
		\\
		&H\Phi_{n,\pm}^{(i)}(y)=m_n^2\Phi_{n,\pm}^{(i)}(y)\,,
		\quad 
		Q\Phi_{n,\pm}^{(i)}(y)=m_n\Phi_{n,\mp}^{(i)}(y)\,, 
		\quad 
		(-1)^F\Phi_{n,\pm}^{(i)}(y)=\pm\Phi_{n,\pm}^{(i)}(y)\,.
	\end{align}
Then we can regard the operators $H$, $Q$, $(-1)^{F}$ and the functions $\Phi_{n,\pm}^{(i)}(y)$ as the Hamiltonian, supercharge, fermion number operator, and bosonic and  fermionic states in the supersymmetric quantum mechanics, respectively.
In the supersymmetric quantum mechanics, it is known that there is a topological invariant which is called the Witten index $\Delta_{\mathrm{W}}$. This index is defined by the difference of the number of the zero energy states with  the eigenvalue $(-1)^F=+1$ and with $(-1)^F=-1$. The topological property of this index is due to the fact that nonzero energy states with $(-1)^F=+1$ and $(-1)^F=-1$ should be paired with each other by the supercharge $Q$, and can move to or from zero energy states together.
In our model, this index corresponds to 
\begin{align}
	\Delta_{\mathrm{W}}= N_{f_{0}}-N_{g_{0}} =N-k,
\end{align}
and then the number of chiral zero modes  becomes a topological invariant.
We will see that this index plays important roles for the topological classification of the boundary conditions with symmetries in Section \ref{sec:Index in each symmetry classes}.

\section{Tenfold classification of boundary conditions with symmetries}
\label{sec:Tenfold classification of boundary conditions with symmetries}

So far, we have not considered symmetries except for the 4d Lorentz invariance.
It should be emphasized that even if the 5d Dirac equation or action
is invariant under some transformations, our model does not necessarily
have those symmetries since transformed fields may not
always satisfy the boundary conditions. Therefore, in order for our
model to have the symmetries, additional restrictions should be on the boundary
conditions at the vertex.

Here, we introduce the time-reversal and charge conjugation symmetries combined with some extra-spatial symmetries and  show that the boundary matrix $U_{\mathrm{B}}$ can be classified into ten classes by those symmetries.
We reveal these classes correspond to the ones in the  classification of SPT phases of zero-dimensional noninteracting gapped  fermions with the  Altland--Zirnbauer (AZ) 
symmetries, which gives the tenfold classification of topological insulators and superconductors.
In this section, we will discuss the correspondence between the boundary matrix $U_{\mathrm{B}}$ and the zero-dimensional Hamiltonian for the gapped free-fermion system, and also show that the symmetries in our model provide the restrictions for $U_{\mathrm{B}}$ identical to the ones for the Hamiltonian from the AZ symmetries.  The correspondence of topological properties will be given in Section \ref{sec:Index in each symmetry classes}.

\subsection{Topological classification of gapped free-fermion system}
\label{subsec:Topological classification of gapped free-fermion system}

The topological insulators and superconductors of fully gapped free-fermion systems  are  topologically classified with AZ symmetries which denote three nonspatial discrete symmetries: time-reversal
symmetry (TRS), particle-hole symmetry (PHS) and chiral symmetry (CS), i.e.  a single-particle Hamiltonian is classified with the presence or absence of the following  symmetries:
  \begin{alignat}{2}
  \text{TRS}
   & :TH({\bf k})T^{-1}=H(-\bf k)&\qquad&(T^{2}=\pm1)\,,
   \label{eq:TRS-0}\\
  \text{PHS}
   & :CH({\bf k})C^{-1}=-H(-\bf k)&\qquad&(C^{2}=\pm1)\,,
   \label{eq:PHS-0}\\
  \text{CS}
   & :\Gamma H({\bf k})\Gamma^{-1}=-H({\bf k})&\qquad&(\Gamma^{2}=1)\,,
   \label{eq:CS-0}
  \end{alignat}
where $H({\bf k})$ is the Hamiltonian in the momentum space with the momentum ${\bf k}$ and  we assume that there are no nontrivial unitary symmetries which commute with the Hamiltonian. If there are such symmetries, we take the Hamiltonian as a block diagonal form and treat each irreducible block as the Hamiltonian $H$ in (\ref{eq:TRS-0})--(\ref{eq:CS-0}).
The TRS and PHS are the antiunitary transformations whose squares should be $+1$ or $-1$. 
The CS is the unitary transformation and can be given by the product  $\Gamma=TC$ up to a phase. Here we take the CS as its square to be $+1$. This symmetry is always present when both the TRS and PHS are present, although the CS can be symmetry alone even if the TRS and PHS are absent.
It should be noted that we can assume there are single TRS, PHS and CS. 
For example, if there are two TRS such as $T_1$ and $T_2$, we can consider the unitary symmetry $T_1T_2$ which commutes with the  Hamiltonian. Then the Hamiltonian can be taken to the block diagonal form and $T_1$ and $T_2$ are trivially related in each irreducible block, which is regarded as $H$ in (\ref{eq:TRS-0})--(\ref{eq:CS-0}).

 It is known that we can obtain total ten symmetry classes with topological numbers as shown in Table \ref{table:AZ_symmetry_class}~\cite{Schnyder:2008tya,Kitaev:2009mg,Ryu:2010zza}.
These ten symmetry classes were originally introduced by A. Altland
and M. R. Zirnbauer and are therefore called AZ
symmetry classes \cite{Zirnbauer:1996zz,Altland:1997zz}.
The topological numbers $\mathbb{Z}$, $\mathbb{Z}_2$ and $2\mathbb{Z}$ in Table \ref{table:AZ_symmetry_class} indicate the presence of nontrivial topological insulators or superconductors. This topological classification is obtained by classifying symmetry-allowed mass terms in the Dirac Hamiltonian and the topological numbers correspond to the 0-th homotopy groups of the classifying spaces, which is the parameter space of the symmetry-allowed mass terms with taking the limit of an infinite number of  energy bands. See Table \ref{table:classifying_space}. These topological numbers characterize whether Hamiltonians can continuously deform to each other or not without closing a mass gap or breaking the symmetries. 
When they belong to the same symmetry class and have the same topological number, they can be continuously deformed to each other. One of the significant features of the topological insulators and superconductors is that 
if there are boundaries in the system and the bulk Hamiltonians in each side have different topological numbers, topologically protected gapless states appear on the boundaries. This is well known as the bulk-boundary correspondence.

\begin{table}[H]
	\captionsetup{width=.90\linewidth}
	\caption{Tenfold classification of topological insulators and superconductors. The class A,\ AIII,\ AI,\ BDI,\ $\cdots$ represent the ten AZ symmetry classes of the Hamiltonian. The signs $+1$ in the chiral symmetry $\Gamma$ and $\pm1$ in the time-reversal $T$ and the particle-hole $C$ mean the presence of those symmetries and also they denote the squares of corresponding symmetries, while $0$ indicates the absence of the symmetries. 
	$d$ is the spatial dimension of the system and $V_d$ denotes the classifying space in each class for the $d$ dimensional space given in Table \ref{table:classifying_space}. $\mathbb{Z}$, $\mathbb{Z}_2$, $2\mathbb{Z}$ and $0$ in the other entries mean the presence or absence of nontrivial topological insulators or superconductors. The class A and AIII are called the complex AZ symmetry classes and have twofold periodicity, while the other classes are referred to the real AZ symmetry classes and have eightfold periodicity. These periodicities are known as the Bott periodicity, which are related to the structure of the K-theory and Clifford algebra.}
	\label{table:AZ_symmetry_class}
	\centering
	\small
	\begin{tabular}{lccccccccccccc}
	\hline
	Class & $T$ & $C$ & $\Gamma$ & $V_d$ & $\pi_{0}(V_{0})$  & $\pi_{0}(V_{1})$ & $\pi_{0}(V_{2})$ & $\pi_{0}(V_{3})$ & $\pi_{0}(V_{4})$ & $\pi_{0}(V_{5})$ & $\pi_{0}(V_{6})$ & $\pi_{0}(V_{7})$ 
	\tabularnewline
	\hline
	\hline
	A & 0 & 0 & 0 & $C_{0+d}$ & $\mathbb{Z}$ & 0 & $\mathbb{Z}$ & 0 & $\mathbb{Z}$ & 0 & $\mathbb{Z}$ & 0 
	\tabularnewline
	AIII & 0 & 0 & 1 & $C_{1+d}$ & 0 & $\mathbb{Z}$ & 0 & $\mathbb{Z}$ & 0 & $\mathbb{Z}$ & 0 & $\mathbb{Z}$
	\tabularnewline
	\hline
	AI & $+1$ & 0 & 0 & $R_{0+d}$ & $\mathbb{Z}$ & 0  & 0 & 0 & 2$\mathbb{Z}$ & 0 & $\mathbb{Z}_2$ & $\mathbb{Z}_2$
	\tabularnewline
	BDI & $+1$ & $+1$ & 1 & $R_{1+d}$ & $\mathbb{Z}_{2}$ & $\mathbb{Z}$ & 0  & 0 & 0 & 2$\mathbb{Z}$ & 0 & $\mathbb{Z}_2$ 
	\tabularnewline
	D & 0 & $+1$ & 0 & $R_{2+d}$ & $\mathbb{Z}_{2}$ & $\mathbb{Z}_{2}$ & $\mathbb{Z}$ & 0  & 0 & 0 & 2$\mathbb{Z}$ & 0 
	\tabularnewline
	DIII & $-1$ & $+1$ & 1 & $R_{3+d}$ & 0 & $\mathbb{Z}_{2}$ & $\mathbb{Z}_{2}$ & $\mathbb{Z}$ & 0  & 0 & 0 & 2$\mathbb{Z}$
	\tabularnewline
	AII & $-1$ & 0 & 0 & $R_{4+d}$ & $2\mathbb{Z}$ & 0 & $\mathbb{Z}_{2}$ & $\mathbb{Z}_{2}$ & $\mathbb{Z}$ & 0  & 0 & 0
	\tabularnewline
	CII & $-1$ & $-1$ & 1 & $R_{5+d}$ & 0 & $2\mathbb{Z}$ & 0 & $\mathbb{Z}_{2}$ & $\mathbb{Z}_{2}$ & $\mathbb{Z}$ & 0  & 0
	\tabularnewline
	C & 0 & $-1$ & 0 & $R_{6+d}$ & 0 & 0 & $2\mathbb{Z}$ & 0 & $\mathbb{Z}_{2}$ & $\mathbb{Z}_{2}$ & $\mathbb{Z}$ & 0
	\tabularnewline
	CI & $+1$ & $-1$ & 1 & $R_{7+d}$ & 0 & 0 & 0 & $2\mathbb{Z}$ & 0 & $\mathbb{Z}_{2}$ & $\mathbb{Z}_{2}$ & $\mathbb{Z}$
	\tabularnewline
	\hline
\end{tabular}
\end{table}
\vspace{1cm}
\begin{table}[H]
	\captionsetup{width=.90\linewidth}
	\caption{Classifying spaces and 0-th homotopy groups. The integers $p$ and $q$ are related to the number of empty bands and occupied bands, respectively. Here, taking the limit of $p$ means that there are an infinite number of empty bands and we focus on the stable classification which is independent of the addition of empty bands. From the viewpoint of the K-theory, this limit results from the property of the so-called stable equivalence.}
	\label{table:classifying_space}
	\centering
	\begin{minipage}[t]{.48\textwidth}
		\small
		\begin{center}
			\begin{tabular}{clc}
				\hline
				$\ell$ mod 2 & Complex classifying space $C_\ell$ & $\pi_{0}(C_\ell)$
				\tabularnewline
				\hline
				\hline
				$\ell=0$ & $C_{0}={\displaystyle \bigcup_{q}\lim_{p\to\infty}\frac{U(p+q)}{U(p)\times U(q)}}$ & $\mathbb{Z}$
				\tabularnewline
				$\ell=1$ & $C_{1}={\displaystyle \lim_{p\to\infty}U(p)}$ & 0
				\tabularnewline
				\hline
			\end{tabular}
		\end{center}
	\end{minipage}
	\hfill
	\begin{minipage}[t]{.48\textwidth}
		\small
		\begin{center}
				\begin{tabular}{clc}
					\hline
					$\ell$ mod 8 & Real classifying space $R_\ell$ & $\pi_{0}(R_\ell)$
					\tabularnewline
					\hline
					\hline
					$\ell=0$ & $R_{0}={\displaystyle \bigcup_{q}\lim_{p\to\infty}\frac{O(p+q)}{O(p)\times O(q)}}$ & $\mathbb{Z}$
					\tabularnewline
					$\ell=1$ & $R_{1}={\displaystyle \lim_{p\to\infty}O(p)}$ & $\mathbb{Z}_{2}$
					\tabularnewline
					$\ell=2$ & $R_{2}={\displaystyle \lim_{p\to\infty}\frac{O(2p)}{U(p)}}$ & $\mathbb{Z}_{2}$
					\tabularnewline
					$\ell=3$ & $R_{3}={\displaystyle \lim_{p\to\infty}\frac{U(2p)}{\it Sp(p)}}$ & 0
					\tabularnewline
					$\ell=4$ & $R_{4}={\displaystyle \bigcup_{q}\lim_{p\to\infty}\frac{{\it Sp}(p+q)}{{\it Sp}(p)\times {\it Sp}(q)}}$ & $2\mathbb{Z}$
					\tabularnewline
					$\ell=5$ & $R_{5}={\displaystyle \lim_{p\to\infty}{\it Sp}(p)}$ & 0
					\tabularnewline
					$\ell=6$ & $R_{6}={\displaystyle \lim_{p\to\infty}\frac{{\it Sp}(p)}{U(p)}}$ & 0
					\tabularnewline
					$\ell=7$ & $R_{7}={\displaystyle \lim_{p\to\infty}\frac{U(p)}{O(p)}}$ & 0\tabularnewline
					\hline
				\end{tabular}
		\end{center}
	\end{minipage}
\end{table}
\vspace{1cm}

\subsection{Correspondence to zero-dimensional Hamiltonian}

Let us focus on the zero-dimensional Hamiltonian. In this case, the Hamiltonian has no momentum dependence. Therefore it consists of only a mass term and is given by a constant Hermitian matrix.  As long as the mass gap does not close and symmetries are restored, the topological structure does not change. Then we can discuss the topological classification by deforming  the Hamiltonian to the following normalized form without closing the gap:
\begin{align}
	H^2=1\,,\ \ \ \ \ \ H^\dagger=H\,.
\end{align}
This Hamiltonian is called a flattened Hamiltonian and has eigenvalues $\pm1$.
We can aware that this condition is  the same as  Eq.\:\eqref{eq:U2-condition} and there is a correspondence between the flattened Hamiltonian  and  the boundary matrix $U_{\mathrm{B}}$. Then, it is expected that the classification of the zero-dimensional topological insulators and superconductors can be applied to that of the boundary conditions in our model. 

To show the correspondence of the classification, we need symmetries which correspond to the TRS and PHS in the zero-dimensional gapped free Hamiltonian.
In the following subsections, we introduce the time-reversal and charge conjugation symmetries combined with some extra-spatial symmetries in our model and show that these provide restrictions for the boundary matrix $U_{\mathrm{B}}$ consistent with the conditions (\ref{eq:TRS-0})--(\ref{eq:CS-0}).

\subsection{Transformations in the $y$-direction}
First of all, we introduce transformations that act only in the extra
dimensional direction, i.e. $y$-direction. These transformations
will be used later when we construct the symmetries corresponding
to the AZ symmetry classes. For convenience, we will label the edge
$L_{a-1}<y<L_{a}$ on the rose graph as
  \begin{align}
    D_{a}=\{y\mid L_{a-1}<y<L_{a}\}\,,\quad a=1,2,\cdots,N\,.
  \end{align}

\subsubsection{Permutation $S_{y}$}

\begin{figure}[h]
	\centering
	\includegraphics[height=6cm]{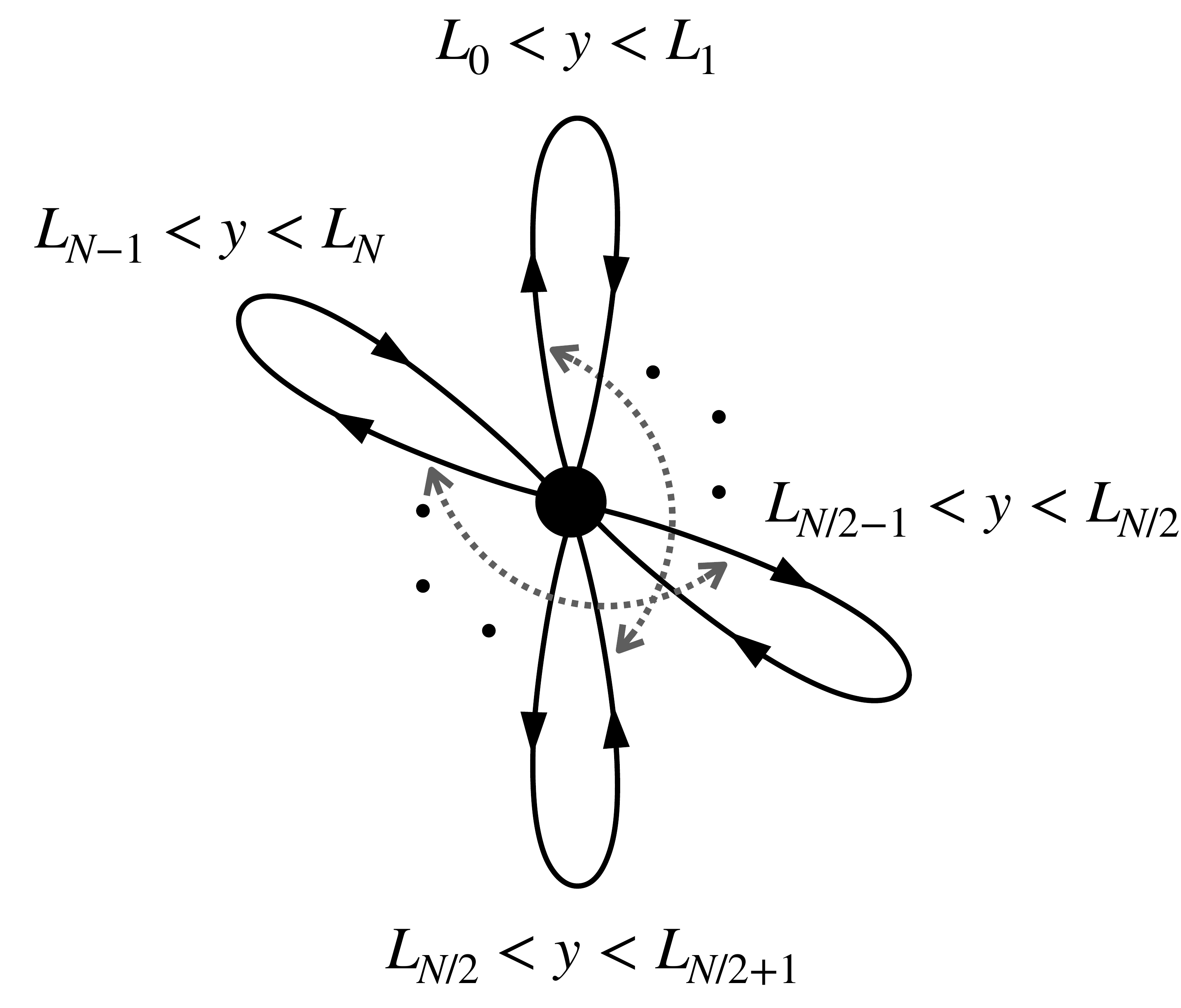}
	\captionsetup{width=.90\linewidth}
	\caption{Permutation $S_{y}$ that exchanges the 
		$a$-th edge for the $(N/2+a)$-th edge $(a=1,\cdots,N/2)$.}
	\label{fig:transformation_1}
\end{figure}

We introduce a permutation $S_y$ that exchanges the 
$a$-edge to the $(N/2+a)$-edge $(a=1,\cdots,N/2)$. This permutation is well-defined
when the $a$-edge and the $(N/2+a)$-edge have the same length
  \begin{align}
     & L_{a}-L_{a-1}=L_{a+N/2}-L_{a-1+N/2}\quad(a=1,2,\cdots,N/2)
  \end{align}
and the number of edges $N$ is even. 
This transformation can be also regarded as the translation of the $a$-edge to the $(N/2+a)$-edge. The mode functions $\varphi(y)=\{f_{n}^{(i)}(y)\:\text{or}\:g_{n}^{(i)}(y)\}$
on $D_{a}$ are transformed as
  \begin{align}
    (S_{y}\varphi)(y)=
    \begin{cases}
    \varphi(y+L_{a-1+N/2}-L_{a-1})& (y\in D_{a}\,,\ a=1,2,\cdots,N/2)\,,
    \\
    \varphi(y+L_{a-1-N/2}-L_{a-1})\,,& (y\in D_{a}\,,\ a=N/2+1,\cdots,N)\,.
    \end{cases}
  \end{align}
Figure \ref{fig:transformation_1} is a diagram of the transformation
$S_{y}$. 

Under the permutation $S_{y}$, the boundary vectors $\vec{F}_{n}^{(i)}$ and
$\vec{G}_{m}^{(j)}$ transform as follows:
  \begin{align}
    \vec{F}_{n}^{(i)} & \xrightarrow{S_{y}}
    \begin{pmatrix}
      f_{n}^{(i)}(L_{N/2}+\varepsilon)\\
      f_{n}^{(i)}(L_{N/2+1}-\varepsilon)\\
      \vdots\\
      f_{n}^{(i)}(L_{N-1}+\varepsilon)\\
      f_{n}^{(i)}(L_{N}-\varepsilon)\\
      \hline f_{n}^{(i)}(L_{0}+\varepsilon)\\
      f_{n}^{(i)}(L_{1}-\varepsilon)\\
      \vdots\\
      f_{n}^{(i)}(L_{N/2-1}+\varepsilon)\\
      f_{n}^{(i)}(L_{N/2}-\varepsilon)
    \end{pmatrix}
    =\begin{pmatrix}
      0 & 1_{N}\\
      1_{N} & 0
    \end{pmatrix}
    \begin{pmatrix}
      f_{n}^{(i)}(L_{0}+\varepsilon)\\
      f_{n}^{(i)}(L_{1}-\varepsilon)\\
      \vdots\\
      f_{n}^{(i)}(L_{N/2-1}+\varepsilon)\\
      f_{n}^{(i)}(L_{N/2}-\varepsilon)\\
      \hline f_{n}^{(i)}(L_{N/2}+\varepsilon)\\
      f_{n}^{(i)}(L_{N/2+1}-\varepsilon)\\
      \vdots\\
      f_{n}^{(i)}(L_{N-1}+\varepsilon)\\
      f_{n}^{(i)}(L_{N}-\varepsilon)
    \end{pmatrix}
    =(\sigma_{1}\otimes1_{N})\vec{F}_{n}^{(i)}\,,
    \label{eq:Ty-1.1}\\
    \vec{G}_{m}^{(j)} & \xrightarrow{S_{y}}
    \begin{pmatrix}
      g_{m}^{(j)}(L_{N/2}+\varepsilon)\\
      -g_{m}^{(j)}(L_{N/2+1}-\varepsilon)\\
      \vdots\\
      g_{m}^{(j)}(L_{N-1}+\varepsilon)\\
      -g_{m}^{(j)}(L_{N}-\varepsilon)\\
      \hline g_{m}^{(j)}(L_{0}+\varepsilon)\\
      -g_{m}^{(j)}(L_{1}-\varepsilon)\\
      \vdots\\
      g_{m}^{(j)}(L_{N/2-1}+\varepsilon)\\
      -g_{m}^{(j)}(L_{N/2}-\varepsilon)
    \end{pmatrix}
    =\begin{pmatrix}
      0 & 1_{N}\\
      1_{N} & 0
    \end{pmatrix}
    \begin{pmatrix}
      g_{m}^{(j)}(L_{0}+\varepsilon)\\
      -g_{m}^{(j)}(L_{1}-\varepsilon)\\
      \vdots\\
      g_{m}^{(j)}(L_{N/2-1}+\varepsilon)\\
      -g_{m}^{(j)}(L_{N/2}-\varepsilon)\\
      \hline g_{m}^{(j)}(L_{N/2}+\varepsilon)\\
      -g_{m}^{(j)}(L_{N/2+1}-\varepsilon)\\
      \vdots\\
      g_{m}^{(j)}(L_{N-1}+\varepsilon)\\
      -g_{m}^{(j)}(L_{N}-\varepsilon)
    \end{pmatrix}
    =(\sigma_{1}\otimes1_{N})\vec{G}_{m}^{(j)}\,.
    \label{eq:Ty-1.2}
  \end{align}

\subsubsection{Half-reflection $R_{y}$}

\begin{figure}[h]
	\centering
	\includegraphics[height=6cm]{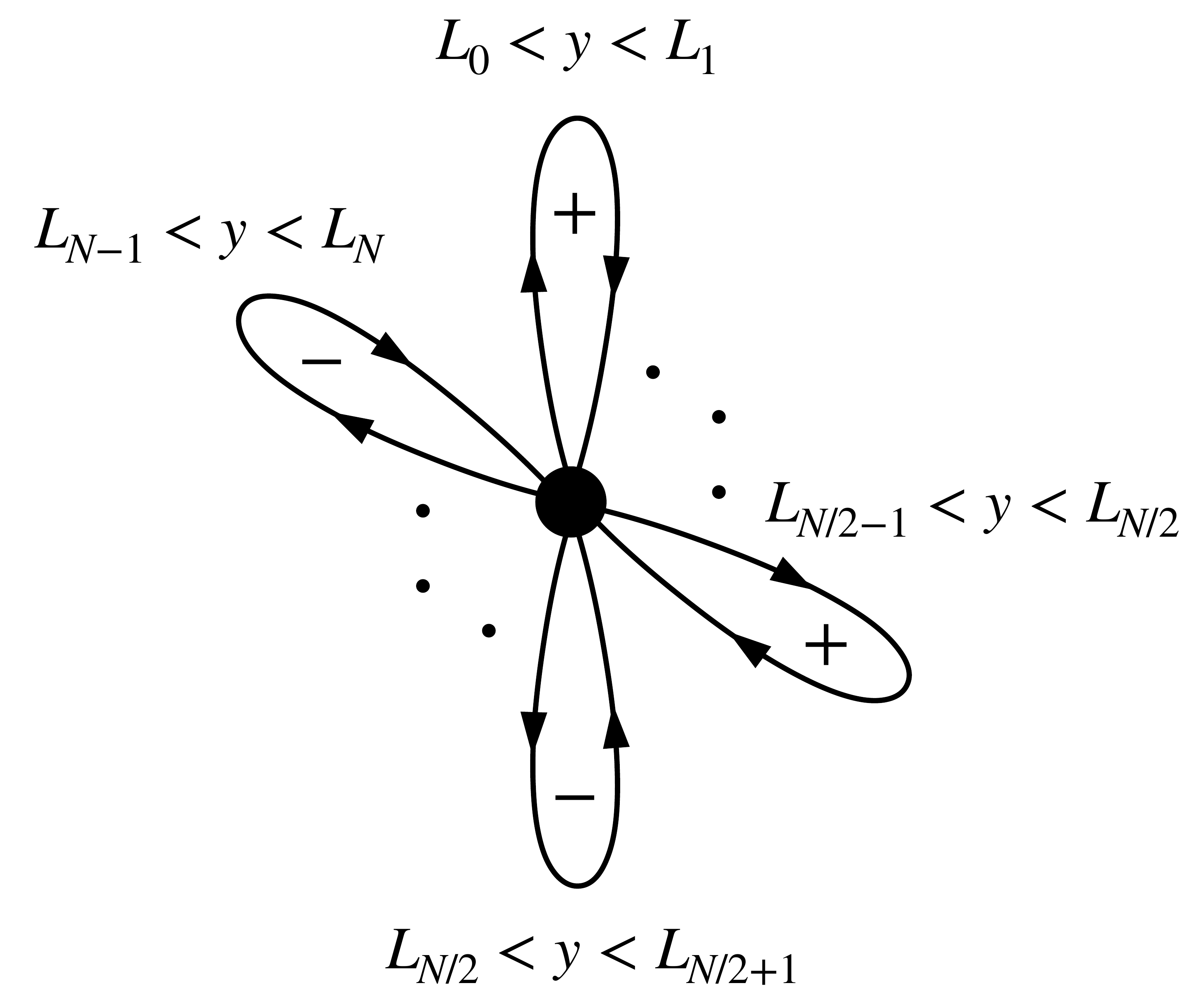}
	\captionsetup{width=.90\linewidth}
	\caption{Half-reflection $R_{y}$ that multiplies the mode
		functions on $D_{a}\:(a=N/2+1,\cdots,2N)$ by $-1$.}
	\label{fig:transformation_2}
\end{figure}

Next, we consider a transformation $R_{y}$ that multiplies the mode function on
$D_{a}\:(a=N/2+1,\cdots,2N)$ by $-1$. This
is called the half-reflection conversion because only signs of the mode functions on the half sections $D_{a}\:(a=N/2+1,\cdots,2N)$
are flipped as if they were reflected
in a mirror. This transformation is well-defined when the quantum
graph has even edges. It should be emphasized that the transformation $R_y$ on the mode functions does not change the mass eigenvalues.
Figure \ref{fig:transformation_2} is a diagram of half-reflection
$R_{y}$. 

The mode functions $\varphi(y)=\{f_{n}^{(i)}(y)\:\text{or}\:g_{n}^{(i)}(y)\}$
on $D_{a}\:(a=1,2,\cdots,N)$ are transformed as
  \begin{align}
    (R_{y}\varphi)(y)=
      \begin{cases}
        \varphi(y) & (y\in D_{a}\,,\ a=1,2,\cdots,N/2)\,,\\
        -\varphi(y) & (y\in D_{a}\,,\ a=N/2+1,\cdots,N)\,.
      \end{cases}
  \end{align}
Then, this half-reflection $R_{y}$ acts on the boundary vectors $\vec{F}_{n}^{(i)}$
and $\vec{G}_{m}^{(j)}$ as 
  \begin{align}
  \vec{F}_{n}^{(i)}
   & \xrightarrow{R_{y}}(\sigma_{3}\otimes1_{N})\vec{F}_{n}^{(i)}\,,
   \label{eq:Ry-1.1}\\
  \vec{G}_{m}^{(j)}
   & \xrightarrow{R_{y}}(\sigma_{3}\otimes1_{N})\vec{G}_{m}^{(j)}\,.
   \label{eq:Ry-1.2}
  \end{align}

\subsubsection{Composite transformation $Q_{y}=-iR_{y}S_{y}$}
Furthermore, we can consider the transformation $Q_{y}$ which is given by the product of $S_{y}$ and $R_{y}$ 
  \begin{align}
    Q_{y}=-iR_{y}S_{y}\,,
  \end{align}
where we suppose that $S_{y}$ acts on the mode function
first, and then $R_{y}$ acts on it.\footnote{It is worth noting that operators $Q_{y}$, $R_{y}$ and $S_{y}$ form
the $SU(2)$ algebra and that the operators $(\mathcal{O}_{1},\mathcal{O}_{2},\mathcal{O}_{3})=(Q_{y},R_{y},S_{y})$ satisfy the 
following relations :
  \[
    \{\mathcal{O}_{i},\mathcal{O}_{j}\}=2\delta_{ij},\quad[\mathcal{O}_{i},\mathcal{O}_{j}]=2i\varepsilon_{ijk}\mathcal{O}_{k}\quad(i,j=1,2,3)\,.
  \]
}

The mode functions $\varphi(y)=\{f_{n}^{(i)}(y)\:\text{or}\:g_{n}^{(i)}(y)\}$
on $D_{a}\:(a=1,2,\cdots,N)$ are transformed as
  \begin{align}
    (Q_{y}\varphi)(y)=
      \begin{cases}
        -i\varphi(y+L_{a-1+N/2}-L_{a-1}) & (y\in D_{a},\ \ a=1,\cdots,N/2)\,,\\
        i\varphi(y+L_{a-1-N/2}-L_{a-1}) & (y\in D_{a},\ \ a=N/2+1,\cdots,N)\,,
      \end{cases}
  \end{align} 
and the transformation for the boundary vectors $\vec{F}_{n}^{(i)}$ and
$\vec{G}_{m}^{(j)}$ are given as
  \begin{align}
    \vec{F}_{n}^{(i)}
     & \xrightarrow{Q_{y}}(\sigma_{2}\otimes1_{N})\vec{F}_{n}^{(i)}\,,
    \label{eq:Qy-1}\\
    \vec{G}_{m}^{(j)}
     & \xrightarrow{Q_{y}}(\sigma_{2}\otimes1_{N})\vec{G}_{m}^{(j)}\,.
    \label{eq:Qy-2}
  \end{align}

\subsubsection{Parity in the $y$-direction $P_{y}$}
\begin{figure}[h]
	\centering
	\includegraphics[height=6cm]{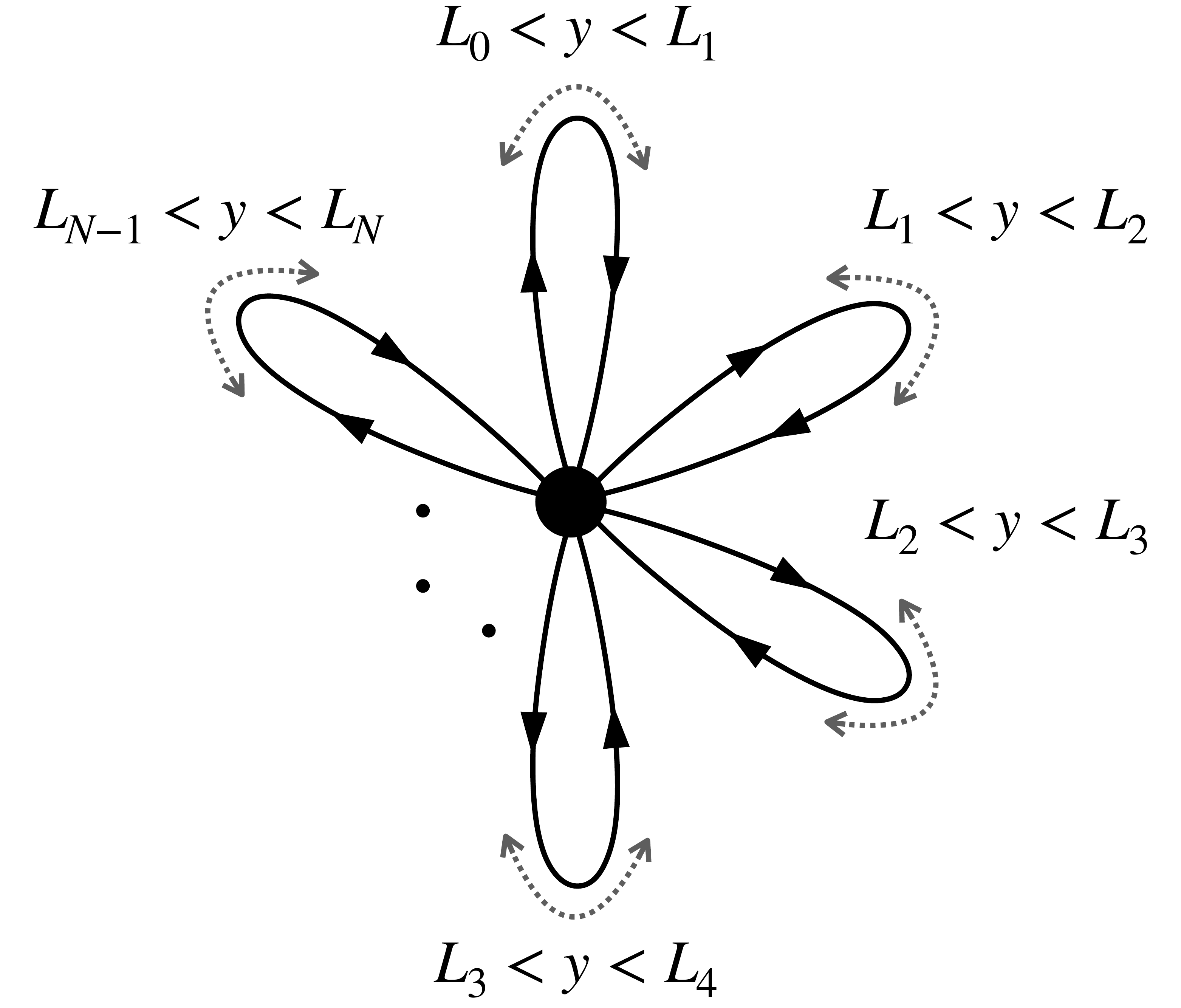}
	\captionsetup{width=.90\linewidth}
	\caption{Parity in the $y$-direction $P_{y}$ that inverts
		the coordinates in each edge.}
	
	\label{fig:transformation_3}
\end{figure}

Finally, we introduce the parity $P_{y}$ that inverts the coordinates in each
edge described in Figure \ref{fig:transformation_3}. The 
transformation for the mode functions $\varphi(y)=\{f_{n}^{(i)}(y)\ \text{or}\ g_{n}^{(i)}(y)\}$
on $D_{a}\:(a=1,2,\cdots,N)$  is given by
  \begin{align}
    &(P_{y}\varphi)(y)=\varphi(L_{a}-y+L_{a-1}) \qquad y\in D_{a}\:(a=1,\cdots,N)\,,
   \end{align}
and the boundary vectors transform as 
	\begin{align}
    &\vec{F}_{n}^{(i)}
     \xrightarrow{P_{y}}(1_{N}\otimes\sigma_{1})\vec{F}_{n}^{(i)}\,,
     \\
    &\vec{G}_{m}^{(j)}
     \xrightarrow{P_{y}}-(1_{N}\otimes\sigma_{1})\vec{G}_{m}^{(j)}\,.
  \end{align}

\subsection{Correspondence to time-reversal symmetries}
Here we define the two types of time-reversal symmetries in the 5d spacetime, the one of which is combined with the extra-spatial transformation $Q_y$.
These time-reversal symmetries lead to the restrictions for the boundary matrix $U_{\mathrm{B}}$. We show that these restrictions correspond to the condition of the TRS \eqref{eq:TRS-0} in the AZ symmetry classes.

\subsubsection{Time-reversal $\mathcal{T}_{+}$}
First, we consider the ordinary time-reversal transformation $\mathcal{T}_{+}$ for the 5d Dirac
fermion
  \begin{align}
    \Psi(x,y)\xrightarrow{\mathcal{T}_{+}}\Psi^{\mathcal{T}_{+}}(x,y)=U_{T}\Psi^{*}(-x^{0},x^{i},y)\,,
    \label{eq:T-transformation-2}
  \end{align}
where $U_{T}$ is defined as a $4\times4$ unitary matrix that satisfies
the following relation:\footnote{In the chiral representation $\gamma^0=\sigma_1\otimes 1_2\,,\  \gamma^i=-i\sigma_2\otimes \sigma_i \,,\ \gamma^y=-i\sigma_3\otimes 1_2$,
the matrix $U_T$ is given by $U_T=\gamma^1\gamma^3$ up to a phase.
}
  \begin{align}
    U_{T}(\gamma^{A})^{*}U_{T}^{-1}=
      \begin{cases}
        \gamma^{0} & (A=0)\,,\\
        -\gamma^{i} & (A=i=1,2,3)\,,\\
        -\gamma^{y} & (A=y)\,.
      \end{cases}
    \label{eq:T-transformation-2.1}
  \end{align}
Although the 5d Dirac equation is invariant under this transformation, we should further require that the boundary conditions (\ref{eq:BC-F}) and (\ref{eq:BC-G}) hold even after the transformation (\ref{eq:T-transformation-2}) in order that $\mathcal{T}_{+}$ becomes a symmetry in our model.

Then, let us substitute the KK decomposition \eqref{eq:2-KK_expantion1} into \eqref{eq:T-transformation-2} to derive the restrictions for $U_{\mathrm{B}}$:
  \begin{align}
     & \sum_{i}\sum_{n}\left[\psi_{\mathrm{R},n}^{(i)}(x)f_{n}^{(i)}(y)+\psi_{\mathrm{L},n}^{(i)}(x)g_{n}^{(i)}(y)\right]\nonumber \\
     & \xrightarrow{\mathcal{T}_{+}}\sum_{i}\sum_{n}U_{T}\psi_{\mathrm{R},n}^{(i)*}(-x^{0},x^{i})f_{n}^{(i)*}(y)+\sum_{i}\sum_{n}U_{T}\psi_{\mathrm{L},n}^{(i)*}(-x^{0},x^{i})g_{n}^{(i)*}(y)\,.
  \end{align}
Taking account of the chirality in four dimensions, we obtain the transformations 
for the 4d fields $\psi_{\mathrm{R/L},n}^{(i)}$ and the mode functions $f_{n}^{(i)},\:g_{n}^{(i)}$ as
  \begin{align}
    \psi_{\mathrm{R/L},n}^{(i)}(x)\xrightarrow{\mathcal{T}_{+}}U_{T}\psi_{\mathrm{R/L},n}^{(i)*}(-x^{0},x^{i})\,,
    \quad 
    f_{n}^{(i)}(y)\xrightarrow{\mathcal{T}_{+}}f_{n}^{(i)*}(y)\,,
    \quad g_{n}^{(i)}(y)\xrightarrow{\mathcal{T}_{+}}g_{n}^{(i)*}(y)\,.
    \label{eq:T-transformation-3}
  \end{align}
The 4d part follows the usual 4d Dirac equation and gives no restrictions.
 On the other hand, the following additional relations must hold for the transformation of the mode functions:
  \begin{align} 
    &(1_{2N}-U_{\mathrm{B}})\vec{F}_{n}^{(i)*}=0\,,
    \\
    &(1_{2N}+U_{\mathrm{B}})\vec{G}_{m}^{(j)*}=0\,.
  \end{align}

Comparing these relations with the complex conjugate of
the original boundary conditions (\ref{eq:BC-F}) and (\ref{eq:BC-G}),
we obtain the restriction
  \begin{align}
    T_{+}U_{\mathrm{B}}T_{+}^{-1}=U_{\mathrm{B}},\quad T_{+}\equiv\mathcal{K}\,,
    \label{eq:T-transformation-4}
  \end{align}
where $\mathcal{K}$ is a complex conjugate operator that acts like
$\mathcal{K}z\mathcal{K}^{-1}=z^{*}$ on the complex number $z$.
Since $T_{+}$ is antiunitary and satisfies
  \begin{align}
    T_{+}^{2}=1, 
    \label{eq:T_+_square}
  \end{align}
Eq.\:\eqref{eq:T-transformation-4} implies that $T_{+}$ corresponds to the TRS with $T^{2}=1$ in the AZ symmetry classes.

\subsubsection{Time-reversal $\mathcal{T}_{-}$}
Next, let us consider a symmetry which leads to the restriction corresponds to the TRS with $T^{2}=-1$ in the AZ symmetry classes.
Here we introduce a  transformation that combines $Q_{y}$ with the
transformation of $\mathcal{T}_{+}$
  \begin{align}
    \Psi(x,y)\xrightarrow{\mathcal{T}_{-}}\Psi^{\mathcal{T}_{-}}(x,y)=Q_{y}U_{T}\Psi^*(-x^{0},x^{i},y)\,,
    \label{eq:T-transformation-5}
  \end{align}
where  $U_{T}$ is defined by Eq.\:(\ref{eq:T-transformation-2.1}).
The only difference from the case of $\mathcal{T}_{+}$ is that the transformation
$Q_{y}$  is included and we will see that this plays the role to flip the sign of the square \eqref{eq:T_+_square}.
$Q_{y}$  does not affect the 5D Dirac equation described on each edge and then the 5d Dirac equation is also invariant for the transformation \eqref{eq:T-transformation-5}. 
However, the transformed field should satisfy the same boundary condition as the original one in order for $\mathcal{T}_{-}$ to be a symmetry. 

If we compare the 4d
chirality in the same way, we obtain the transformation $\mathcal{T}_{+}$ for the 4d fields and the mode functions as
  \begin{align}
    \psi_{\mathrm{R/L},n}^{(i)}(x)\xrightarrow{\mathcal{T}_{-}}U_{T}\psi_{\mathrm{R/L},n}^{(i)*}(-x^{0},x^{i})\,,\quad f_{n}^{(i)}(y)\xrightarrow{\mathcal{T}_{-}}Q_{y}f_{n}^{(i)*}(y)\,,\quad g_{n}^{(i)}(y)\xrightarrow{\mathcal{T}_{-}}Q_{y}g_{n}^{(i)*}(y)\,.
    \label{eq:T-transformation-6}
  \end{align}
Then, in order for our model to have the time-reversal symmetry $\mathcal{T}_{-}$, we find that the relations 
  \begin{align}
    (1_{2N}-U_{\mathrm{B}})(\sigma_{2}\otimes1_{N})\vec{F}_{n}^{(i)*}
     & =0\,,\\
    (1_{2N}+U_{\mathrm{B}})(\sigma_{2}\otimes1_{N})\vec{G}_{m}^{(j)*}
     & =0
  \end{align}
must hold from Eqs.\: (\ref{eq:Qy-1}) and (\ref{eq:Qy-2}).

Comparing these relations with the complex conjugate of the original boundary
condition (\ref{eq:BC-F}) and (\ref{eq:BC-G}), we obtain the restriction
  \begin{align}
    T_{-}U_{\mathrm{B}}T_{-}^{-1}=U_{\mathrm{B}}\,,\quad T_{-}\equiv(i\sigma_{2}\otimes1_{N})\mathcal{K}\,.
    \label{eq:T-transformation-7}
  \end{align}
$T_{-}$ is antiunitary and its square is 
  \begin{align}
    T_{-}^{2}=-1.
  \end{align}
Therefore Eq.\:\eqref{eq:T-transformation-7} implies $T_{-}$ corresponds to the
TRS with $T^{2}=-1$ in the AZ symmetry classes.

\subsection{Correspondence to particle-hole symmetries}
We consider two types of charge conjugations with extra-spatial transformations similar to the time-reversal symmetries and show those provide the restrictions for $U_{\mathrm{B}}$. They correspond to the PHS \eqref{eq:PHS-0} in the AZ symmetry classes.

\subsubsection{Charge conjugation $\mathcal{C}_{-}$}
First, we introduce a transformation defined by the 4d charge conjugation with the parity $P_y$ in the extra space which is consistent with the 4d Lorentz symmetry in our model:
  \begin{align}
    \Psi(x,y)\xrightarrow{\mathcal{C}_{-}}\Psi^{\mathcal{C}_{-}}(x,y)=P_{y}U_{C}\overline{\Psi}^{\top}(x,y)\,,
    \label{eq:C-transformation-1}
  \end{align}
where $U_{C}$ is the usual 4d charge conjugation matrix defined by\footnote{In the chiral representation $\gamma^0=\sigma_1\otimes 1_2\,,\  \gamma^i=-i\sigma_2\otimes \sigma_i \,,\ \gamma^y=-i\sigma_3\otimes 1_2$,
	the matrix $U_C$ is given by $U_C=\gamma^0\gamma^2$ up to a phase.
}
  \begin{align}
    U_{C}(\gamma^{\mu})^{\top}U_{C}^{-1}=-\gamma^{\mu}\,,\quad U_{C}^{\top}=-U_{C}\,.
    \label{eq:C-transformation-1.1}
  \end{align}
From the above relation between $U_{C}$ and $\gamma^{\mu}$,
the gamma matrix $\gamma^{y}\:(=-i\gamma^{5})$ satisfies
  \begin{align}
    U_{C}(\gamma^{y})^{\top}U_{C}^{-1}=\gamma^{y}\,.
    \label{eq:C-transformation-1.2}
  \end{align}
In this paper, we refer to this transformation as the charge conjugation $\mathcal{C}_{-}$.
While the 5d Dirac equation is not invariant by only the 4d charge conjugation due to the sign of the right-hand side in Eq.\:\eqref{eq:C-transformation-1.2}, it is invariant by $\mathcal{C}_{-}$ since this transformation additionally includes the parity $P_y$.\footnote{Although we focus on the $\mathcal{C}_{-}$ symmetry in this paper, we can also consider a 5d charge conjugation without the parity which is given by
\begin{align*}
		&\Psi(x,y)\rightarrow U_{C'}\overline{\Psi}^{\top}(x,y)\,, \quad U_{C'}^{\top}=-U_{C'}\,, \quad U_{C'}(\gamma^{A})^{\top}U_{C'}^{-1}=+\gamma^{A}\ \ \ (A=0,\cdots,3,y)\,.
\end{align*}
The bulk mass should vanish under this symmetry unlike the case of $\mathcal{C}_{-}$. See \cite{Kugo:1982bn} for detail.
}

In addition to the invariance of the 5d Dirac equation, the field after the transformation should satisfy the boundary condition in order for $\mathcal{C}_{-}$ to be a symmetry.
By substituting the KK decomposition (\ref{eq:2-KK_expantion1})
into (\ref{eq:C-transformation-1})
and comparing the chirality in four dimensions, we obtain the transformation $\mathcal{C}_{-}$ for the 4d fields and the mode functions as
  \begin{align}
    \psi_{\mathrm{R/L},n}^{(i)}(x)\xrightarrow{\mathcal{C}_{-}}U_{C}\overline{\psi_{\mathrm{L/R},n}^{(i)}}^{\top}(x)\,,\quad f_{n}^{(i)}(y)\xrightarrow{\mathcal{C}_{-}}P_{y}g_{n}^{(i)*}(y)\,,\quad g_{n}^{(i)}(y)\xrightarrow{\mathcal{C}_{-}}P_{y}f_{n}^{(i)*}(y)\,.
    \label{eq:C-transformation-2}
  \end{align}
The mode functions $f_n^{(i)}$ and $g_n^{(i)}$ are interchanged because of the change of the 4d chirality.
Therefore, if there exist zero mode functions $f^{(i)}_0$, zero modes $g^{(i)}_0$ should also exist and we can take it as 
	\begin{align}
			g^{(i)}_0(y)=P_yf^{(i)*}_0(y)\,.
			\label{eq:C-transformation-zero_mode}
	\end{align}
For massive modes $(n\neq0)$,  since the relations between $f^{(i)}_n(y)$ and $g^{(i)}_n(y)$ are already fixed by Eqs.\:\eqref{eq:2-SUSYrelations1} and \eqref{eq:2-SUSYrelations2}, $P_yg^{(i)*}_n(y)\ (P_yf^{(i)*}_n(y))$ is given by  linear combinations of $f^{(i)}_n(y)\ (g^{(i)}_n(y))$.

The transformation for the boundary vectors $\vec{F}_{n}^{(i)}$ and $\vec{G}_{m}^{(j)}$
are 
  \begin{align}
    \vec{F}_{n}^{(i)}
     & \xrightarrow{\mathcal{C}_{-}}-(1_{N}\otimes i\sigma_{2})\vec{G}_{n}^{(i)*}\,,
     \label{eq:C-transforamtion-2.3}\\
    \vec{G}_{m}^{(j)}
     & \xrightarrow{\mathcal{C}_{-}}(1_{N}\otimes i\sigma_{2})\vec{F}_{m}^{(j)*}\,,
     \label{eq:C-transforamtion-2.4}
   \end{align}
and  the following relations must hold:
  \begin{align}
    &(1_{2N}-U_{\mathrm{B}})(1_{N}\otimes i\sigma_{2})\vec{G}_{n}^{(i)*}=0\,,
    \\
    &(1_{2N}+U_{\mathrm{B}})(1_{N}\otimes i\sigma_{2})\vec{F}_{m}^{(j)*}=0\,.
  \end{align}

From the the original boundary conditions (\ref{eq:BC-F}) and (\ref{eq:BC-G}), the above relations give the restriction for $U_{\mathrm{B}}$
  \begin{align}
    C_{-}U_{\mathrm{B}}C_{-}^{-1}=-U_{\mathrm{B}}\,,\quad C_{-}\equiv(1_{N}\otimes i\sigma_{2})\mathcal{K}\,.
    \label{eq:C-transformation-3}
  \end{align}
Here $C_{-}$ is antiunitary and satisfies
  \begin{align}
    C_{-}^{2}=-1\,.
  \end{align}
Then we can find that $C_{-}$ corresponds to the
PHS with $C^{2}=-1$ in the AZ symmetry classes.

\subsubsection{Charge conjugation $\mathcal{C}_{+}$}
\label{subsubsec:C+}
Next, we consider the transformation $\mathcal{C}_{+}$ which consists of the charge conjugation $\mathcal{C}_{-}$ with the transformation $Q_y$
  \begin{align}
     & \Psi(x,y)\xrightarrow{\mathcal{C}_{+}}\Psi^{\mathcal{C}_{+}}(x,y)=Q_{y}P_{y}U_{C}\overline{\Psi}^{\top}(x,y)\,,
     \label{eq:C-transformation-4}
  \end{align}
where $U_{C}$ is defined by Eq.\:(\ref{eq:C-transformation-1.1}). The 5d Dirac equation is also invariant by this transformation.

Then let us consider the restriction for $U_{\mathrm{B}}$ in order for $\mathcal{C}_{+}$ to be a symmetry.
The transformation for the 4d fields and the mode functions are given by
  \begin{align}
    \psi_{\mathrm{R/L},n}^{(i)}(x)\xrightarrow{\mathcal{C}_{+}}U_{C}\overline{\psi_{\mathrm{L/R},n}^{(i)}}^{\top}(x)\,,\quad f_{n}^{(i)}(y)\xrightarrow{\mathcal{C}_{+}}Q_{y}P_{y}g_{n}^{(i)*}(y),\quad g_{n}^{(i)}(y)\xrightarrow{\mathcal{C}_{+}}Q_{y}P_{y}f_{n}^{(i)*}(y)\,.
    \label{eq:C-transformation-5}
  \end{align}
Then the boundary vectors $\vec{F}_{n}^{(i)}$ and $\vec{G}_{m}^{(j)}$
are transformed as
  \begin{align}
    \vec{F}_{n}^{(i)}\xrightarrow{\mathcal{C}_{+}}
     & -(\sigma_{2}\otimes1_{N/2}\otimes i\sigma_{2})\vec{G}_{n}^{(i)*}\,,\\
    \vec{G}_{m}^{(j)}\xrightarrow{\mathcal{C}_{+}}
     & (\sigma_{2}\otimes1_{N/2}\otimes i\sigma_{2})\vec{F}_{m}^{(j)*}\,.
  \end{align}
Therefore
the following relations must hold:
  \begin{align}
    (1_{2N}-U_{\mathrm{B}})(\sigma_{2}\otimes1_{N/2}\otimes i\sigma_{2})\vec{G}_{n}^{(i)*}
    & =0\,,\\
    (1_{2N}+U_{\mathrm{B}})(\sigma_{2}\otimes1_{N/2}\otimes i\sigma_{2})\vec{F}_{m}^{(j)*}
    & =0\,.
  \end{align}

Comparing these relations with
the original boundary conditions (\ref{eq:BC-F}) and (\ref{eq:BC-G}),
we obtain the restriction
  \begin{align}
    C_{+}U_{\mathrm{B}}C_{+}^{-1}=-U_{\mathrm{B}},\quad C_{+}\equiv(i\sigma_{2}\otimes1_{N/2}\otimes i\sigma_{2})\mathcal{K}\,.
    \label{eq:C-transformation-6}
  \end{align}
$C_{+}$ is antiunitary and its square is 
  \begin{align}
    C_{+}^{2}=+1.
  \end{align}
Therefore  $C_{+}$ corresponds to the
PHS with $C^{2}=+1$ in the AZ symmetry classes.

\subsection{Correspondence to chiral symmetry}
Finally, let us discuss the symmetries which are obtained by the product of the time-reversal and charge conjugation transformations discussed above. We show that these symmetries lead to restrictions for $U_{\mathrm{B}}$ and  correspond to the CS \eqref{eq:CS-0} in the AZ symmetry classes.

\subsubsection{Chiral symmetry $\Gamma_{+}$}
We introduce the transformation of the product $\mathcal{T}_{+}\mathcal{C}_{+}$  or  $\mathcal{T}_{-}\mathcal{C}_{-}$.
From Eqs.\:(\ref{eq:T-transformation-2}), (\ref{eq:C-transformation-4}) or 
(\ref{eq:T-transformation-5}), (\ref{eq:C-transformation-1}),
the transformation properties of $\Psi(x,y)$ are given by
  \begin{align}
    \Psi(x,y)
    \xrightarrow{\mathcal{T}_{+}\mathcal{C}_{+}}-Q_{y}P_{y} U_{T}U_{C}^{*}\gamma^{0}\Psi(-x^{0},x^{i},y)\,,
    \label{eq:P-transformation-2.1}
    \\
     \Psi(x,y)
    \xrightarrow{\mathcal{T}_{-}\mathcal{C}_{-}}+Q_{y}P_{y} U_{T}U_{C}^{*}\gamma^{0}\Psi(-x^{0},x^{i},y)\,.
    \label{eq:P-transformation-2.2}
  \end{align}
Thus, these transformations are equivalent up to the sign.
Comparing the chirality in four dimensions, we obtain the transformations for the 4d fields and the mode functions as  
  \begin{align}
    \psi_{\mathrm{R/L},n}^{(i)}(x)&\xrightarrow{\mathcal{T}_{\pm}\mathcal{C}_{\pm}}U_{T}U_{C}^{*}\gamma^{0}\psi_{\mathrm{L/R},n}^{(i)}(-x^{0},x^{i})\,,
    \\
    f_{n}^{(i)}(y)&\xrightarrow{\mathcal{T}_{\pm}\mathcal{C}_{\pm}}\mp Q_{y}P_{y}g_{n}^{(i)}(y)\,,\quad g_{n}^{(i)}(y)\xrightarrow{\mathcal{T}_{\pm}\mathcal{C}_{\pm}}\mp Q_{y}P_{y}f_{n}^{(i)}(y)\,.
    \label{eq:C-transformation-2.2}
  \end{align}
Therefore, the boundary vectors $\vec{F}_{n}^{(i)}$ and $\vec{G}_{m}^{(j)}$
are transformed as
  \begin{align}
    \vec{F}_{n}^{(i)}
    & \xrightarrow{\mathcal{T}_{\pm}\mathcal{C}_{\pm}}\pm(\sigma_{2}\otimes1_{N/2}\otimes i\sigma_{2})\vec{G}_{n}^{(i)}\,,\\
    \vec{G}_{m}^{(j)}
    & \xrightarrow{\mathcal{T}_{\pm}\mathcal{C}_{\pm}}\mp(\sigma_{2}\otimes1_{N/2}\otimes i\sigma_{2})\vec{F}_{m}^{(j)}\,.
  \end{align}
These indicate that the relations
  \begin{align}
    (1_{2N}-U_{\mathrm{B}})(\sigma_{2}\otimes1_{N/2}\otimes i\sigma_{2})\vec{G}_{n}^{(i)}
    & =0\,,\\
    (1_{2N}+U_{\mathrm{B}})(\sigma_{2}\otimes1_{N/2}\otimes i\sigma_{2})\vec{F}_{n}^{(i)}
    & =0
  \end{align}
must hold in order for $\mathcal{T}_{\pm}\mathcal{C}_{\pm}$ to be a symmetry.
We then obtain the restriction for $U_{\mathrm{B}}$
  \begin{align}
    \Gamma_{+}U_{\mathrm{B}}\Gamma_{+}^{-1}=-U_{\mathrm{B}},\quad\Gamma_{+}\equiv i\sigma_{2}\otimes1_{N/2}\otimes i\sigma_{2}\,.
  \end{align}
$\Gamma_{+}$ is unitary and we can find that  $\Gamma_{+}$ corresponds to the
CS in the AZ symmetry classes.
In
terms of operators that act on $U_{\mathrm{B}}$,
we can confirm the  relation
  \begin{align}
    \Gamma_+=T_\pm C_\pm\,.
  \end{align}

\subsubsection{Chiral symmetry $\Gamma_{-}$}
We can consider the another transformation of the product
$\mathcal{T}_{+}\mathcal{C}_{-}$ or  equivalently $\mathcal{T}_{-}\mathcal{C}_{+}$
  \begin{align}
    \Psi(x,y)
    & \xrightarrow{\mathcal{T}_{\pm}\mathcal{C}_{\mp}}\pm P_{y}U_{T}U_{C}^{*}\gamma^0\Psi(-x^{0},x^{i},y)\,.
  \end{align}
The transformation for the 4d fields and the mode functions are then given by
  \begin{align}
    \psi_{\mathrm{R/L},n}^{(i)}(x)&\xrightarrow{\mathcal{T}_{\pm}\mathcal{C}_{\mp}}U_{T}U_{C}^{*}\gamma^0\psi_{\mathrm{L/R},n}^{(i)}(-x^{0},x^{i})\,,
    \\
     f_{n}^{(i)}(y)&\xrightarrow{\mathcal{T}_{\pm}\mathcal{C}_{\mp}}\pm(P_{y}g_{n}^{(i)})(y),\quad g_{n}^{(i)}(y)\xrightarrow{\mathcal{T}_{\pm}\mathcal{C}_{\mp}}\pm(P_{y}f_{n}^{(i)})(y)\,.
    \label{eq:C-transformation-3.2}
  \end{align}
Therefore, the boundary vectors $\vec{F}_{n}^{(i)}$ and $\vec{G}_{m}^{(j)}$
are transformed as
  \begin{align}
    \vec{F}_{n}^{(i)} & \xrightarrow{\mathcal{T}_{\pm}\mathcal{C}_{\mp}}\mp(1_{N}\otimes i\sigma_{2})\vec{G}_{n}^{(i)}\,,
    \label{eq:C-transforamtion-2.3-1-1}\\
    \vec{G}_{m}^{(j)} & \xrightarrow{\mathcal{T}_{\pm}\mathcal{C}_{\mp}}\pm(1_{N}\otimes i\sigma_{2})\vec{F}_{m}^{(j)}\,,
    \label{eq:C-transforamtion-2.4-1-1}
  \end{align}
and the following relations should be satisfied in order for $\mathcal{T}_{\pm}\mathcal{C}_{\mp}$ to be a symmetry:
  \begin{align}
    (1_{2N}-U_{\mathrm{B}})(1_{N}\otimes i\sigma_{2})\vec{G}_{n}^{(i)}
    & =0\,,\\
    (1_{2N}+U_{\mathrm{B}})(1_{N}\otimes i\sigma_{2})\vec{F}_{n}^{(i)}
    & =0\,.
  \end{align}

These relations yield the restriction for $U_{\mathrm{B}}$
  \begin{align}
    \Gamma_{-}U_{\mathrm{B}}\Gamma_{-}^{-1}=-U_{\mathrm{B}}\,,\quad\Gamma_{-}\equiv1_{N}\otimes\sigma_{2}\,.
  \end{align}
$\Gamma_{-}$ is unitary and we can find that  $\Gamma_{-}$ corresponds to the
chiral symmetry in the AZ symmetry classes.
We can also confirm that $\Gamma_{-}$ can be given by
  \begin{align}
  	\Gamma_-=\mp iT_\pm C_\mp\,.
  \end{align}

\subsection{Summary of correspondence to AZ symmetry class}
In this section, we considered the time-reversal and the charge conjugation with the extra-spatial transformations in our model and it was revealed that these symmetries provide the restrictions for the  boundary matrix $U_{\mathrm{B}}$  as shown in Table~\ref{table:correspondence_symmetry}. These correspond to the TRS, PHS and CS in the AZ symmetry classes.
The matrix $U_{\mathrm{B}}$ corresponds to the zero-dimensional Hamiltonian $H$ in (\ref{eq:TRS-0})--(\ref{eq:CS-0}) and therefore we can classify $U_{\mathrm{B}}$ into ten symmetry classes as in Table~\ref{table:AZ_symmetry_class} for the $d=0$ case.

It should be noted that we assume only one of the same type symmetries such as $\mathcal{T}_+$ and $\mathcal{T}_-$ can be present so far.
If both symmetries are present, $Q_y$ also becomes the symmetry independently since $Q_y$ can be given by their product.
In this case, we consider the  identification of the $(a+N/2)$-edge to $a$-edge ($a=1,\cdots,N/2$) by the symmetry $Q_y$ with the eigenvalues $Q_y=+1$ or $Q_y=-1$, and then classify the boundary conditions of this reduced system.  This identification effectively reduces the rose graph with $N$ edges to the one with $N/2$ edges like the $S^1$ is reduced to the interval by the $\mathbb{Z}_2$ orbifold, and the same type symmetries such as $\mathcal{T}_+$ and $\mathcal{T}_-$ are trivially related with each other after the identification. 
The symmetry $Q_y$ requires that $U_{\mathrm{B}}$ commutes with $\sigma_2\otimes1_N$ from  Eqs.\:\eqref{eq:Qy-1} and \eqref{eq:Qy-2}, and the matrix $U_{\mathrm{B}}$ can be written as 
				\begin{align}
				U_B=\frac{1_{2}+\sigma_2}{2}\otimes u_{\mathrm{B}+}+\frac{1_{2}-\sigma_2}{2}\otimes u_{\mathrm{B}-}\,,
			\end{align}
 where $u_{\mathrm{B}\pm}$ are $N\times N$ Hermitian unitary matrices.  This $u_{\mathrm{B}+}\ (u_{\mathrm{B}-})$ specifies the boundary condition for  the reduced rose graph with $N/2$ edges with $Q_y=+1\ (Q_y=-1)$ and corresponds to the irreducible blocks in the Hamiltonian for the gapped free-fermion system discussed in Section \ref{subsec:Topological classification of gapped free-fermion system}.

\begin{table}[H]
	\centering
	\captionsetup{width=.90\linewidth}
	\caption{The transformations in our model and  the correspondence to the AZ symmetries.}
	\label{table:correspondence_symmetry}
  \begin{tabular}{lll}
    \hline
    AZ symmetry & Transformation for $\Psi(x,y)$ & Restriction for $U_{\mathrm{B}}$
    \tabularnewline
    \hline
    Time-reversal  & $\Psi(x,y)\xrightarrow{\mathcal{T}_{+}}U_{T}\mathcal{K}\Psi(-x^{0},x^{i},y)$ & $T_{+}U_{\mathrm{B}}T_{+}^{-1}=U_{\mathrm{B}}$\tabularnewline
    \ \ \ \ \ $(T^2=+1)$&&$T_{+}=\mathcal{K}$
    \vspace{0.3cm}\tabularnewline
    Time-reversal  & $\Psi(x,y)\xrightarrow{\mathcal{T}_{-}}Q_{y}U_{T}\mathcal{K}\Psi(-x^{0},x^{i},y)$ & $T_{-}U_{\mathrm{B}}T_{-}^{-1}=U_{\mathrm{B}}$\tabularnewline
    \ \ \ \ \ $(T^2=-1)$&&$T_{-}=(i\sigma_{2}\otimes1_{N})\mathcal{K}$
    \vspace{0.3cm}\tabularnewline
    \hline
    Particle-hole & $\Psi(x,y)\xrightarrow{\mathcal{C}_{+}}Q_{y}P_{y}U_{C}\overline{\Psi}^{\top}(x,y)$ & $C_{+}U_{\mathrm{B}}C_{+}^{-1}=-U_{\mathrm{B}}$\tabularnewline
    \ \ \ \ \ $(C^2=+1)$&&$C_{+}=(i\sigma_{2}\otimes1_{N/2}\otimes i\sigma_{2})\mathcal{K}$
    \vspace{0.3cm}\tabularnewline
    Particle-hole  & $\Psi(x,y)\xrightarrow{\mathcal{C}_{-}}P_{y}U_{C}\overline{\Psi}^{\top}(x,y)$ & $C_{-}U_{\mathrm{B}}C_{-}^{-1}=-U_{\mathrm{B}}$\tabularnewline
    \ \ \ \ \ $(C^2=-1)$&&$C_{-}=(1_{N}\otimes i\sigma_{2})\mathcal{K}$
    \vspace{0.3cm}\tabularnewline
    \hline
    Chiral  & $\Psi(x,y)\xrightarrow{\mathcal{T}_{\pm}\mathcal{C}_{\pm}}\mp Q_{y}P_{y}U_{T}U_{C}^{*}\gamma^{0}\Psi(-x^{0},x^{i},y)$ & $\Gamma_{+}U_{\mathrm{B}}\Gamma_{+}^{-1}=-U_{\mathrm{B}}$\tabularnewline
    &&$\Gamma_{+}=i\sigma_{2}\otimes1_{N/2}\otimes i\sigma_{2}$
    \vspace{0.3cm}\tabularnewline
    Chiral & $\Psi(x,y)\xrightarrow{\mathcal{T}_{\pm}\mathcal{C}_{\mp}}\pm P_{y}U_{T}U_{C}^{*}\gamma^{0}\Psi(-x^{0},x^{i},y)$ & $\Gamma_{-}U_{\mathrm{B}}\Gamma_{-}^{-1}=-U_{\mathrm{B}}$\tabularnewline
     &&$\Gamma_{-}=1_{N}\otimes\sigma_{2}$
     \vspace{0.3cm}\tabularnewline
    \hline
  \end{tabular}
\par\end{table}

\section{Index and zero modes in each symmetry class}
\label{sec:Index in each symmetry classes}
From the correspondence of the boundary conditions in our model and the zero-dimensional gapped free-fermion system, we can obtain the nontrivial topological numbers $\mathbb{Z}, \mathbb{Z}_2$ and $2\mathbb{Z}$ for the boundary conditions in each symmetry class as well as the topological insulators and superconductors in Table \ref{table:AZ_symmetry_class} for the $d=0$ case.
In the topological matter side, the topological numbers specify the number of gapless states which appear on boundaries.
Then, the question is what  do these topological numbers physically mean in our model?

In this section, we will reveal that these topological numbers correspond to the numbers of zero modes localized at the vertex in our model as summarized in Table \ref{table:our-result}.
The topological numbers $\mathbb{Z}$ and $2\mathbb{Z}$ are related to the  Witten index, which describes the number of chiral zero modes given in Section \ref{subsec:Witten index}. By considering a sufficiently large number of the edges $N$, this number can take any integer values in the class A and AI and also any multiple of two in the class AII.
The large $N$ limit corresponds to taking an infinite number of bands in the zero-dimensional Hamiltonian.
In addition, the topological number $\mathbb{Z}_2$ in the class BDI and D corresponds to the number of Dirac zero modes in module 2. Here we call it $\mathbb{Z}_2$ index.
We will see that the $\mathbb{Z}_2$ index becomes topological invariant due to the additional degeneracy of the massive modes by the symmetry $\mathcal{C}_+$.
We also investigate the classifying spaces of $U_{\mathrm{B}}$ in our model, which are the parameter spaces of $U_{\mathrm{B}}$  restricted by symmetry conditions, and show  they are identical to the ones in the gapped free-fermion system.

\begin{table}[H]
	\captionsetup{width=.90\linewidth}
	\caption{Tenfold classification of the boundary conditions in our model. The sign $\pm 1$ in the column of $T$ and $C$ denote the presence of $T_\pm$ and $C_\pm$, respectively and  also 1 in $\Gamma$ indicates the presence of the chiral symmetry $\Gamma_+$ or $\Gamma_-$, while 0 means the absence of corresponding symmetries.
	We also describe the Witten index  for the type  $(2N-k,k)$ BC in each symmetry class, which is equivalent to the number of the chiral zero modes in  our model. In addition, $\mathbb{Z}_2$ in the column of the $\mathbb{Z}_2$ index indicates that the number of massless 4d Dirac fields in module 2 is topologically nontrivial, while 0 means topologically trivial and the number of massless 4d Dirac fields can be zero by continuous deformations of  parameters.
	These correspond to the topological numbers in Table \ref{table:AZ_symmetry_class} for the $d=0$ case.
}
	\label{table:our-result}
	\centering{}%
	\begin{tabular}{lccclccc}
		\hline
		\multirow{2}{*}{Class} & \multirow{2}{*}{$T$} & \multirow{2}{*}{$C$} & \multirow{2}{*}{$\Gamma$} & \multirow{2}{*}{Classifying space of $U_{\mathrm{B}}$} & $\Delta_{\mathrm{W}}$ for & \multirow{2}{*}{$\mathbb{Z}_2$ index }
		\tabularnewline
		&&&&&type $(2N-k,k)$ BC&&
		\tabularnewline
		\hline
		\hline
		A & 0 & 0 & 0 & $C_0={\displaystyle \bigcup_{k=0}^{2N}\frac{U(2N)}{U(2N-k)\times U(k)}}$ & $N-k$ & 0 
		\vspace{0.2cm}
		\tabularnewline
		AIII & 0 & 0 & 1 & $C_1=U(N)$ & 0 & 0 
		\vspace{0.2cm}
		\tabularnewline
		\hline
		AI & $+1$ & 0 & 0 & $R_0={\displaystyle \bigcup_{k=0}^{2N}\frac{O(2N)}{O(2N-k)\times O(k)}}$ & $N-k$ & 0 
		\vspace{0.2cm}
		\tabularnewline
		BDI & $+1$ & $+1$ & 1 & $R_1=O(N)$ & 0 & $\mathbb{Z}_{2}$ 
		\vspace{0.2cm}
		\tabularnewline
		D & 0 & $+1$ & 0 & $R_2={\displaystyle \frac{O(2N)}{U(N)}}$ & 0 & $\mathbb{Z}_{2}$ 
		\vspace{0.2cm}
		\tabularnewline
		DIII & $-1$ & $+1$ & 1 & $R_3={\displaystyle \frac{U(N)}{{\it Sp}(N/2)}}$ & 0 & 0 
		\vspace{0.2cm}
		\tabularnewline
		AII & $-1$ & 0 & 0 & $R_4={\displaystyle \bigcup_{k=0}^{2N}\frac{{\it Sp}(N)}{{\it Sp}((2N-k)/2)\times {\it Sp}(k/2)}}$ & $N-k$& 0 
		\vspace{-0.4cm}
		\tabularnewline
		&&&&&$(N,k:\text{even})$
		\vspace{0.2cm}
		\tabularnewline
		CII & $-1$ & $-1$ & 1 & $R_5={\it Sp}(N/2)$ & 0 & 0 
		\vspace{0.2cm}
		\tabularnewline
		C & 0 & $-1$ & 0 & $R_6={\displaystyle \frac{{\it Sp}(N)}{U(N)}}$ & 0 & 0 
		\vspace{0.2cm}
		\tabularnewline
		CI & $+1$ & $-1$ & 1 & $R_7={\displaystyle \frac{U(N)}{O(N)}}$ & 0 & 0 
		\vspace{0.2cm}
		\tabularnewline
		\hline
	\end{tabular}
\end{table}

\subsection{Witten index and classifying spaces}
Here, let us discuss the Witten index and the classifying space for each symmetry class in our model, and show the correspondence to the topological numbers $\mathbb{Z}$, $2\mathbb{Z}$ and the classifying spaces in the gapped free-fermion system.
The $\mathbb{Z}_2$ index will be discussed in the next subsection.

\subsubsection{Class A}
Since the class A has no symmetries, there are no additional conditions
for $U_{\mathrm{B}}$. Therefore, $U_{\mathrm{B}}$ is diagonalized
as follows (see Section \ref{subsec:Boundary conditions}):
  \begin{align}
    U_{\mathrm{B}}=V
      \begin{pmatrix}
        1_{2N-k} & 0\\
        0 & -1_{k}
      \end{pmatrix}
    V^{\dagger}\,,
    \quad V\in U(2N)\quad(k=0,1,\cdots,2N).
  \end{align}
The Witten index is determined by the number of the eigenvalues $\pm1$
of $U_{\mathrm{B}}$, and is given by
  \begin{align}
    \Delta_{\mathrm{W}}=N-k\quad(k=0,1,\cdots,2N)\,.
  \end{align}
By considering a sufficiently large $N$, the Witten index  can take any integer and this corresponds to the topological number $\mathbb{Z}$.

In addition, the parameter space of $U_{\mathrm{B}}$ in the type $(2N-k,k)$ BC is $U(2N)/(U(2N-k)\times U(k))$ from the matrix $V$. Then we can obtain the  classifying space of $U_{\mathrm{B}}$ as 
  \begin{align}
    C_0=\bigcup_{k=0}^{2N}\frac{U(2N)}{U(2N-k)\times U(k)}\,.
  \end{align}
This is also identical to the one for the zero-dimensional Hamiltonian in the gapped free-fermion system.

\subsubsection{Class AIII}
The class AIII has only the CS with the 
unitarity operator $\Gamma$ for $U_{\mathrm{B}}$ which denotes $\Gamma_{+}$ or $\Gamma_{-}$ given in Section \ref{sec:Tenfold classification of boundary conditions with symmetries}. This operator satisfies
  \begin{align}
    \{U_{\mathrm{B}},\Gamma\}=0\,,\quad\Gamma^{2}=1_{2N}\,.
    \label{eq:AIII-1}
  \end{align}
This implies that the boundary vectors of the zero mode functions with the chiral operator $\Gamma\vec{F}_{0}^{(i)}\ (\Gamma\vec{G}_{0}^{(j)})$ satisfy the boundary condition of $\vec{G}_{0}^{(j)}\ (\vec{F}_{0}^{(i)})$.
Therefore,
if the CS is present, $\vec{F}_{0}^{(i)}$ and $\vec{G}_{0}^{(j)}$
have the same degrees of degeneracy, i.e. the equal degrees of freedom for
$i$ and $j$. For this reason, $U_{\mathrm{B}}$ should be diagonalized
as
  \begin{align}
    U_{\mathrm{B}}=V
      \begin{pmatrix}
        1_{N} & 0\\
        0 & -1_{N}
      \end{pmatrix}
    V^{\dagger}\,,
    \quad V\in U(2N)
  \end{align}
and the Witten index is given by
  \begin{align}
    \Delta_{\mathrm{W}}=0\,.
  \end{align}

$\Gamma$ can
be taken to the diagonal form of $\widetilde\Gamma=\sigma_{3}\otimes1_{N}$ by an appropriate basis change such as
	\begin{align}
			U_{\mathrm{B}}&\to\widetilde U_{\mathrm{B}}={\widetilde V}^\dagger U_{\mathrm{B}}{\widetilde V}\,,
			\\
			\Gamma&\to \widetilde\Gamma={\widetilde V}^\dagger\Gamma {\widetilde V}\,,
			\\
			\vec{F}_{n}^{(i)}&\to  {\widetilde V}^\dagger\vec{F}_{n}^{(i)}\,,
			\\
			\vec{G}_{m}^{(j)}&\to  {\widetilde V}^\dagger \vec{G}_{m}^{(j)}\quad {\widetilde V} \in U(2N)\,.
	\end{align}
In this basis,
$\widetilde U_{\mathrm{B}}$ is written as 
  \begin{align}
  \widetilde{U}_{\mathrm{B}}=
    \begin{pmatrix}
      0 & u_{\mathrm{B}}\\
      u_{\mathrm{B}}^{\dagger} & 0
    \end{pmatrix}\,,
    \quad u_{\mathrm{B}}\in U(N)
  \end{align}
from Eq.\:(\ref{eq:AIII-1}) and the conditions ${U}_{\mathrm{B}}^{2}=1_{2N}$ and ${U}_{\mathrm{B}}^\dagger={U}_{\mathrm{B}}$.
Therefore the parameter space of $U_{\mathrm{B}}$ is specified by $u_{\mathrm{B}}$  and then the classifying space is
  \begin{align}
    C_1=U(N)\,.
  \end{align}

\subsubsection{Class AI}
The class AI has the $\mathcal{T}_{+}$
symmetry and the additional condition for  $U_{\mathrm{B}}$ is
  \begin{align}
    T_{+}U_{\mathrm{B}}T_{+}^{-1}=U_{\mathrm{B}}\,,\quad T_{+}=\mathcal{K}\,.
  \end{align}
This requires that $U_{\mathrm{B}}$ is a real
matrix. Then $U_{\mathrm{B}}$ can be written as
  \begin{align}
    U_{\mathrm{B}}=R
      \begin{pmatrix}
        1_{2N-k} & 0\\
        0 & -1_{k}
      \end{pmatrix}
    R^{\top}\,,\quad R\in O(2N)\quad(k=0,1,\cdots,2N)\,.
  \end{align}
Therefore, the Witten index and the classifying space
are given by
  \begin{align}
    \Delta_{\mathrm{W}} & =N-k\quad(k=0,1,\cdots,2N)\,,\\
    R_0& =\bigcup_{k=0}^{2N}\frac{O(2N)}{O(2N-k)\times O(k)}\,,
  \end{align}
respectively. The difference between the class A and AI is that $U_{\mathrm{B}}$
is a real matrix in this class. Therefore the mode functions can be taken to be real since the complex conjugation of the mode functions also satisfy the boundary condition and become solutions of Eqs.\:\eqref{eq:2-eigenequation_1} and \eqref{eq:2-eigenequation_2} with the same mass eigenvalues.

\subsubsection{Class BDI}
\label{subsubsec:Witten index_BDI}
The class BDI has the three symmetries $\mathcal{T}_{+},\ \mathcal{C}_{+}$ and $\mathcal{T}_{+}\mathcal{C}_{+}$. Then $U_{\mathrm{B}}$ satisfies
  \begin{alignat}{2}
    T_{+}U_{\mathrm{B}}T_{+}^{-1}
    & =U_{\mathrm{B}}\,,  \qquad &T_{+}&=\mathcal{K}\,,
    \label{eq:T+_BDI}
    \\
    C_{+}U_{\mathrm{B}}C_{+}^{-1}
    & =-U_{\mathrm{B}}\,,\qquad &C_{+}&=(i\sigma_{2}\otimes1_{N/2}\otimes i\sigma_{2})\mathcal{K}\,,
    \\
    \Gamma_{+}U_{\mathrm{B}}\Gamma_{+}^{-1}
    & =-U_{\mathrm{B}}\,,\qquad &\Gamma_{+}&=i\sigma_{2}\otimes1_{N/2}\otimes i\sigma_{2}\,.
    \label{eq:Gamma_BDI}
  \end{alignat}
From the same discussion in the class AIII, the degeneracy of $\vec{F}_{0}^{(i)}$
and $\vec{G}_{0}^{(j)}$ are equal to each other due to $\Gamma_{+}$ and the Witten index becomes
\begin{align}
	\Delta_{\mathrm{W}} & =0\,.
\end{align}

Then let us discuss the classifying space.
Here we focus on the operators $T_{+}$ and $\Gamma_{+}$ due to the relation $\Gamma_+=T_{+}C_{+}$.
Since $\Gamma_{+}$ is the real symmetric matrix, this can be diagonalized by a real orthogonal matrix.
When we consider a basis change such that
	\begin{align}
			\widetilde\Gamma_{+}&= {\widetilde V}^{\top}\Gamma_{+}{\widetilde V}=\sigma_{3}\otimes1_{N}\,,
			\\
			\widetilde T_{+}&= {\widetilde V}^{\top} T_{+}{\widetilde V}=\mathcal{K}
	\end{align}
with the real orthogonal matrix ${\widetilde V}$
	\begin{align}
			{\widetilde V}=\frac{1}{\sqrt{2}}
			\begin{pmatrix}
				1_N & 1_{N/2}\otimes\sigma_1
				\\
				-1_{N/2}\otimes i\sigma_2 & 1_{N/2}\otimes\sigma_3
			\end{pmatrix}\,.
	\end{align}
$\widetilde U_{\mathrm{B}}={\widetilde V}^{\top} U_{\mathrm{B}}{\widetilde V}$ is given by a real matrix  and restricted to the form
  \begin{align}
    \widetilde{U}_{\mathrm{B}}=\begin{pmatrix}0 & u_{\mathrm{B}}\\
    u_{\mathrm{B}}^{\top} & 0
    \end{pmatrix}\,,\quad 
	u_{\mathrm{B}}\in O(N)
  \end{align} 
by the conditions \eqref{eq:T+_BDI}, \eqref{eq:Gamma_BDI}  and ${U}_{\mathrm{B}}^\dagger={U}_{\mathrm{B}}^{\top}$ and ${U}_{\mathrm{B}}^{\dagger}={U}_{\mathrm{B}}$.
Then we can see that the classifying space of $U_{\mathrm{B}}$
is given by
  \begin{align}
   R_1= O(N)\,.
  \end{align}

\subsubsection{Class D}
\label{subsubsec:Witten index_D}
Since the class D has the $\mathcal{C}_{+}$ symmetry, the condition for $U_{\mathrm{B}}$ is
  \begin{align}
    C_{+}U_{\mathrm{B}}C_{+}^{-1}=-U_{\mathrm{B}}\,,\quad C_{+}=(i\sigma_{2}\otimes1_{N/2}\otimes i\sigma_{2})\mathcal{K}\,.
    \label{eq:C_+_class_D}
  \end{align}
First, let us diagonalize $C_{+}$ as 
  \begin{align}
     & \widetilde C_{+}= {\widetilde V}^\dagger C_{+}{\widetilde V}=\mathcal{K}\,, 
     \label{eq:D-1}
  \end{align}
where
  \begin{align}
  	& {\widetilde V}=\frac{1}{\sqrt{2}}
  	\begin{pmatrix}
  		1_N & 1_{N/2}\otimes i\sigma_1
  		\\
  		-1_{N/2}\otimes i\sigma_2 & 1_{N/2}\otimes i\sigma_3
  	\end{pmatrix}\,.
  \end{align}
In this basis, the condition ${U}_{\mathrm{B}}^\dagger={U}_{\mathrm{B}}$ and Eq.\:\eqref{eq:C_+_class_D} and \eqref{eq:D-1} imply that $\widetilde U_{\mathrm{B}}={\widetilde V}^\dagger U_{\mathrm{B}}{\widetilde V}$ can be of the form
\begin{align}
	\widetilde{U}_{\mathrm{B}}=iA\,,\quad A^{\top}=-A\,,
\end{align}
where $A$ is a $2N\times2N$ antisymmetric real matrix. 
Since any pure imaginary antisymmetric matrix  has the same number of positive and negative eigenvalues, the Witten index is
  \begin{align}
    \Delta_{\mathrm{W}}=0\,.
  \end{align}

Next, let us consider the classifying space in this class.
In general, using a real orthogonal matrix $R$, we can bring the real antisymmetric matrix $A$ into a
block off-diagonal form:
  \begin{align}
    A=R\begin{pmatrix}{0} & 1_{N}\\
    -1_{N} & {0}
    \end{pmatrix}R^{\top}\,,\quad R\in O(2N)\,.
    \label{eq:D-3}
  \end{align}
If there is a matrix $R'$ which satisfies
  \begin{align}
    R^{\prime}
      \begin{pmatrix}
        {0} & 1_{N}\\
        -1_{N} & {0}
      \end{pmatrix}
    R^{\prime\top}=
      \begin{pmatrix}
        {0} & 1_{N}\\
        -1_{N} & {0}
      \end{pmatrix}\,, \quad R'\in {\it SO}(2N)\,,
    \label{eq:D-4}
  \end{align}
the replacement $R\to RR'$  does not change the matrix $A$. Therefore the classifying space is given by the coset space,  which is $O(2N)$ divided by the parameter space of $R'$.
We mention that $R'$ should have the determinant $\det R' =+1$ since this matrix also belongs to the symplectic group.\footnote{From the properties of the Pfaffian, we obtain
\begin{align*}
	\rm{Pf}\left(R' \begin{pmatrix}
		{0} & 1_{N}\\
		-1_{N} & {0}
	\end{pmatrix}
R'^{\top}\right)=\det(R')\cdot \rm{Pf}\begin{pmatrix}
	{0} & 1_{N}\\
	-1_{N} & {0}
\end{pmatrix}\,.
\end{align*}
Substituting Eq.\:\eqref{eq:D-4} into the left-hand side of the above equation, we see that $\det R'=+1$.
}  
We can write $R'$ as
  \begin{align}
    R'=\exp\left\{ 1_{2}\otimes X_{0}+\sigma_{2}\otimes X_{2}\right\} ,
  \end{align}
where $X_{0}$ is an antisymmetric real matrix and $X_{2}$ is a
symmetric pure imaginary matrix. Since $\sigma_{2}$ can be diagonalized
by the unitary matrix
  \begin{align}
    G_{2}=\frac{1}{\sqrt 2}
      \begin{pmatrix}
        1 & -i\\
        1 & i
      \end{pmatrix}
    ,\quad G_{2}^{\dagger}G_{2}=G_{2}G_{2}^{\dagger}=1_{2}\,,
  \end{align}
$R'$ is given by
  \begin{align}
    R'
    & =(G_{2}^{\dagger}\otimes1_{N})\exp\left\{ 1_{2}\otimes X_{0}+\sigma_{3}\otimes X_{2}\right\} (G_{2}\otimes1_{N})\nonumber \\
    & =(G_{2}^{\dagger}\otimes1_{N})
      \begin{pmatrix}
        \exp(X_{0}+X_{2}) & {0}\\
        {0} & \exp(X_{0}-X_{2})
      \end{pmatrix}
    (G_{2}\otimes1_{N})\nonumber \\
    & \equiv(G_{2}^{\dagger}\otimes1_{N})
      \begin{pmatrix}
        r & {0}\\
        {0} & r^{*}
      \end{pmatrix}
    (G_{2}\otimes1_{N})\quad(r\equiv\exp(X_{0}+X_{2}))\,.
  \end{align}
Here $r$ is a unitary matrix because it satisfies $r^{\dagger}r=1_{N}$.
Therefore, it turns out that $R'$ is specified by an element of
$U(N)$.  Then the classifying space of ${U}_{\mathrm{B}}$ is
  \begin{align}
    R_2=\frac{O(2N)}{U(N)}\,.
  \end{align}

\subsubsection{Class DIII}
The class DIII has the three symmetries $\mathcal{T}_-$, $\mathcal{C}_+$ and $\mathcal{T}_-\mathcal{C}_+$.
Then $U_{\mathrm{B}}$ satisfies
  \begin{alignat}{2}
    T_{-}U_{\mathrm{B}}T_{-}^{-1}
    & =U_{\mathrm{B}}\,,
    \qquad &T_{-}&=(i\sigma_{2}\otimes1_{N})\mathcal{K}\,,
    \label{eq:T-_DIII}
    \\
    C_{+}U_{\mathrm{B}}C_{+}^{-1}
    & =-U_{\mathrm{B}}\,,
    \qquad &C_{+}&=(i\sigma_{2}\otimes1_{N/2}\otimes i\sigma_{2})\mathcal{K}\,,\\
    \Gamma_{-}U_{\mathrm{B}}\Gamma_{-}^{-1}
    & =-U_{\mathrm{B}}\,,
    \qquad&\Gamma_{-}&=1_{N}\otimes\sigma_{2}\,.
  \end{alignat}
Since the CS is present, the Witten index in this class should be zero, i.e.
\begin{align}
	\Delta_{\mathrm{W}}=0\,.
\end{align}

$T_{-}$ and $\Gamma_{-}$ are anticommutative and we can take the basis
  \begin{alignat}{2}
  	\widetilde \Gamma_{-}&= {\widetilde V}^\dagger\Gamma_{-}{\widetilde V}=\sigma_{3}\otimes1_{N}\,,
  	\\
    \widetilde T_{-}&={\widetilde V}^\dagger T_{-}{\widetilde V}=
    (\sigma_{1}\otimes i\sigma_{2}\otimes1_{N/2})\mathcal{K}=
      \begin{pmatrix}
        {0} & (i\sigma_{2}\otimes1_{N/2})\mathcal{K}\\
        (i\sigma_{2}\otimes1_{N/2})\mathcal{K} & {0}
      \end{pmatrix}
  \end{alignat}
by the unitary matrix $\widetilde V$
	\begin{align}
			&\widetilde V=W_{N\leftrightarrow 2}\frac{1}{\sqrt 2}
			\begin{pmatrix}
				1_N & 1_N\\
				i1_N & -i1_N
			\end{pmatrix}\,.
	\end{align}
Here we define the real orthogonal matrix $W_{n_a\leftrightarrow n_b}$ 
 which exchanges the order of the direct product of an $n_a\times n_a$ matrix $A$ and an $n_b\times n_b$ matrix $B$ such that
\begin{align}
	&W_{n_a\leftrightarrow n_b}^\top(A\otimes B)W_{n_a\leftrightarrow n_b}=B\otimes A\,,
	\\
	&W_{n_a\leftrightarrow n_b}=\sum_{i,\alpha}(\vec{e}_i\otimes \vec E_\alpha) ( \vec E_\alpha^\top\otimes \vec{e}_i^\top)\,,
	\label{eq:W_exchange}
	\\
	&\vec{e}_i^\top=(\underbrace{0,\cdots,0}_{i-1},1,0,\cdots,0)^{\top}\,,
	\\
&
\vec{E}_{\alpha}^\top=(\underbrace{0,\cdots,0}_{\alpha-1},1,0,\cdots,0)^{\top}
\quad(i=1,\cdots,n_a,\ \ \alpha=1,\cdots,n_b)\,.
\end{align}
In this basis, $\widetilde U_{\mathrm{B}}={\widetilde V}^\dagger U_{\mathrm{B}}{\widetilde V}$ is restricted to the block off-diagonalized
form
  \begin{align}
    \widetilde{U}_{\mathrm{B}}=
      \begin{pmatrix}
        0 & u_{\mathrm{B}}\\
        u_{\mathrm{B}}^{\dagger} & 0
      \end{pmatrix}\,,
  	\quad u_{\mathrm{B}}u_{\mathrm{B}}^{\dagger}=1_{N}\,,
  	\quad(\sigma_{2}\otimes1_{N/2})u_{\mathrm{B}}^{\top}(\sigma_{2}\otimes1_{N/2})=u_{\mathrm{B}}\,.
    \label{eq:DIII-1}
  \end{align}
Since $u_{\mathrm{B}}$ is a unitary matrix, it can be diagonalized
as
  \begin{align}
    u_{\mathrm{B}}=v^{\dagger}u_{\mathrm{d}}v\,,\quad u_{\mathrm{d}}=
      \begin{pmatrix}
        e^{i\alpha_{1}} &  & 0\\
         & \ddots\\
         0 &  & e^{i\alpha_{N}}
      \end{pmatrix}\,,
  \quad v\in U(N)\,,\quad\alpha_{i}\in\mathbb{R}\ \ \ (i=1,\cdots,N)\,.
    \label{eq:DIII-2}
  \end{align}
Substituting the first equation in \eqref{eq:DIII-2} into the third equation in \eqref{eq:DIII-1} with $J_{y}\equiv\sigma_{2}\otimes1_{N/2}$,
we can get the relation
  \begin{align}
    (vJ_{y}v^{\top})u_{\mathrm{d}}=u_{\mathrm{d}}(vJ_{y}v^{\top})\,,
  \end{align}
or equivalently
	\begin{align}
			\quad(vJ_{y}v^{\top})_{ij}(u_{\mathrm{d}})_{jj}=(u_{\mathrm{d}})_{ii}(vJ_{y}v^{\top})_{ij}\qquad(i,j=1,\cdots,N)\,,
	\end{align}
where $i,j$ denote the indices of the matrix elements and are not summed  up.
This implies that $(u_{\mathrm{d}})_{ii}=(u_{\mathrm{d}})_{jj}$ if $(vJ_{y}v^{\top})_{ij}\neq0$. 
Then, by dividing both sides of the above equation by $u_{\mathrm{d}}^{1/2}$,  we can also obtain the following relation:
  \begin{align}
    (vJ_{y}v^{\top})_{ij}\sqrt{(u_{\mathrm{d}})_{jj}}=\sqrt{(u_{\mathrm{d}})_{ii}}(vJ_{y}v^{\top})_{ij}\,.
    \label{eq:DIII-3}
  \end{align}
This relation is also trivially satisfied in the case of $(vJ_{y}v^{\top})_{ij}=0$.
With this relation, $u_{\mathrm{B}}$ can be rewritten as 
\begin{align}
	u_{\mathrm{B}}=w^{\top}J_{y}wJ_{y}\,,
	\qquad w\equiv v^{\top}u_{\mathrm{d}}^{1/2}v^{*}\,.
\end{align}
By definition, $w$ is an $N\times N$ unitary matrix. This $w$ specifies the parameter space of $u_{\mathrm{B}}$.
We can find that $u_{\mathrm{B}}$ is invariant by the transformation
  \begin{align}
  	w\to gw\quad(g^{\dagger}g=1_{N},\quad g^{\top}J_{y}g=J_{y})\,.
  \end{align}
Since $g$ is an element of the symplectic group {\it Sp}$(N/2)$,
the classifying space of $U_{\mathrm{B}}$ is given by
  \begin{align}
    R_3=\frac{U(N)}{{\it Sp}(N/2)}\,.
  \end{align}

\subsubsection{Class AII}
The class AII has the $\mathcal{T}_{-}$ symmetry. The additional condition for $U_{\mathrm{B}}$ is 
  \begin{align}
    T_{-}U_{\mathrm{B}}T_{-}^{-1}=U_{\mathrm{B}}\,,\quad T_{-}=(i\sigma_{2}\otimes1_{N})\mathcal{K}\,.
  \end{align}
Let  $\vec{F}^{(i)}$ and $\vec{G}^{(j)}$ $(i,j=1,2,\cdots)$ are orthonormal eigenvectors with  $U_{\mathrm{B}}=+1$ and $U_{\mathrm{B}}=-1$, respectively.
From the above condition, the vectors $\vec{F}^{(i)}$ and ${T}_{-}\vec{F}^{(i)}$ (with the same index $i$) have the same eigenvalues $U_{\mathrm{B}}=+1$, and they are orthogonal to each other from the direct calculation. 
If we have a boundary vector $\vec{F}^{(j)}\ (i\neq j)$ which is orthogonal with $\vec{F}^{(i)}$ and $T_-\vec{F}^{(i)}$, $T_-\vec{F}^{(j)}$ is also orthogonal with them.
Therefore, the number of the eigenvalue $U_{\mathrm{B}}=+1$ is even in this class.
The same is true for  the case of $\vec{G}^{(j)}$ and ${T}_{-}\vec{G}^{(j)}$, and  the number of the eigenvalue $U_{\mathrm{B}}=-1$ is also even.
Then, in the type $(2N-k,k)$ BC, the number $k$ should be even.
It should be noted that $N$ is also even for the $Q_y$ transformation in $\mathcal{T}_{-}$ to be well-defied.
Thus, the Witten index in this class is given by an even number:
\begin{align}
	\Delta_{\mathrm{W}}=N-k\quad (k=0,2,4,\cdots, N,\ \ N:\text{even})\,.
\end{align}
This corresponds to the topological number $2\mathbb{Z}$.

For the type $(2N-k,k)$ BC,  we can rewrite 
${U}_{\mathrm{B}}$
as
  \begin{align}
    {U}_{\mathrm{B}}=V
      \begin{pmatrix}
      	1_{(2N-k)/2} &  &  & 0\\
         & -1_{k/2}\\
         &  & 1_{(2N-k)/2}\\
        0 &  &  & -1_{k/2}
      \end{pmatrix}V^\dagger\,,
  \end{align}
where
the $2N\times2N$ matrix $V$ is a unitary matrix defined by
\begin{align}
	V=\left(\vec{F}^{(1)},\cdots,\vec{F}^{^{\left(\frac{2N-k}{2}\right)}},\vec{G}^{(1)},\cdots,\vec{G}^{\left(\frac{k}{2}\right)},-{T}_{-}\vec{F}^{(1)},\cdots,-{T}_{-}\vec{F}^{^{\left(\frac{2N-k}{2}\right)}},-{T}_{-}\vec{G}^{(1)},\cdots,-{T}_{-}\vec{G}^{\left(\frac{k}{2}\right)}\right)\,.
	\label{eq:AII-1}
\end{align}
This matrix  is of the form
  \begin{align}
    V=\begin{pmatrix}A & -B^{*}\\
    B & A^{*}
    \end{pmatrix}
    \label{eq:AII-2}
  \end{align}
with $N\times N$ matrices $A$ and
$B$.
Since $V$ is a unitarity matrix, $A$ and $B$ must satisfy
  \begin{align}
    A^{\dagger}A+B^{\dagger}B
    & =1_{N}\,, 
  \\
    -B^{\top}A+A^{\top}B
    & =0\,,
    \\
    AA^\dagger +B^*B^\top&=1_N\,,
    \\
    BA^\dagger-A^*B^\top&=0\,.
  \end{align}
Then $V$ satisfies
  \begin{align}
    V^{\top}
      \begin{pmatrix}
        0 & 1_{N}\\
        -1_{N} & 0
      \end{pmatrix}
    V
    =
    \begin{pmatrix}
      0 & 1_{N}\\
      -1_{N} & 0
    \end{pmatrix}.
    \label{eq:AII-3}
  \end{align}
This implies that the  matrix $V$ is an element of the  symplectic group ${\it Sp}(N)$.
However there is redundancy left in this $V$.
Let 
 \begin{align}
	g_i=\begin{pmatrix}A_i & -B_i^{*}\\
		B_i & A_i^{*}
	\end{pmatrix}
\end{align}
be an element of ${\it Sp}(i)$ and we consider the matrix $g$ which belongs to ${\it Sp}((2N-k)/2)\times {\it Sp}(k/2)$ such as
	\begin{align}
			g=
			\begin{pmatrix}
					A_{(2N-k)/2} & 0 & -B_{(2N-k)/2}^* & 0
					\\
					0 & A_{k/2} & 0 & -B_{k/2}^*
					\\
					B_{(2N-k)/2} & 0 & A_{(2N-k)/2}^* & 0
					\\
					0 & B_{k/2} & 0 & A_{k/2}^*
			\end{pmatrix}\,.
	\end{align}
We find that $U_{\mathrm{B}}$ is invariant by the replacement $V\to Vg$.
Therefore, the classifying space of $U_{\mathrm{B}}$ is given by the coset space
  \begin{align}
    R_4=\bigcup_{k=0}^{N}\frac{{\it Sp}(N)}{{\it Sp}((2N-k)/2)\times {\it Sp}(k/2)}\qquad(N,k:\text{even})\,.
  \end{align}

\subsubsection{Class CII}
The class CII has the three symmetries, $\mathcal{T}_-$, $\mathcal{C}_-$ and $\mathcal{T}_-\mathcal{C}_-$.
Then $U_{\mathrm{B}}$ satisfies
  \begin{alignat}{2}
    T_{-}U_{\mathrm{B}}T_{-}^{-1}
    & =U_{\mathrm{B}},\qquad &T_{-}&=(i\sigma_{2}\otimes1_{N})\mathcal{K}\,,
    \label{eq:T-_CII}\\
    C_{-}U_{\mathrm{B}}C_{-}^{-1}
    & =-U_{\mathrm{B}},\qquad &C_{-}&=(1_{N}\otimes i\sigma_{2})\mathcal{K}\,,\\
    \Gamma_{+}U_{\mathrm{B}}\Gamma_{+}^{-1}
    & =-U_{\mathrm{B}},\qquad&\Gamma_{+}&=i\sigma_{2}\otimes1_{N/2}\otimes i\sigma_{2}\,.
  \end{alignat}
Since the CS is present, the Witten index in the class CII
is
\begin{align}
	\Delta_{\mathrm{W}}=0\,.
\end{align}

When we take the basis 
  \begin{align}
  	\widetilde\Gamma_{+}&= {\widetilde V}^{\top}\Gamma_+{\widetilde V}=\sigma_{3}\otimes1_{N}\,,
  	\\
    \widetilde T_{-}&= {\widetilde V}^{\top} T_{-}{\widetilde V}=(1_{2}\otimes i\sigma_{2}\otimes1_{N/2})\mathcal{K}=
      \begin{pmatrix}
        i\sigma_{2}\otimes1_{N/2} & 0\\
        0 & i\sigma_{2}\otimes1_{N/2}
      \end{pmatrix}
    \mathcal{K}\,,
    \\
    {\widetilde V}&=\frac{1}{\sqrt{2}}
    \begin{pmatrix}
    	1_{N/2}\otimes\sigma_3 & -1_{N/2}\otimes i\sigma_2
    	\\
    	1_{N/2}\otimes \sigma_1 & 1_N
    \end{pmatrix}(1_2\otimes W_{N/2\leftrightarrow2})
  \end{align}
using the matrix $W_{N/2\leftrightarrow2}$ defined by \eqref{eq:W_exchange},
$\widetilde U_{\mathrm{B}}={\widetilde V}^{\top} U_{\mathrm{B}}{\widetilde V}$ is of the form
  \begin{align}
    \widetilde{U}_{\mathrm{B}}=
      \begin{pmatrix}
        0 & u_{\mathrm{B}}\\
        u_{\mathrm{B}}^{\dagger} & 0
      \end{pmatrix}
    \label{eq:CII-1}
  \end{align}
with the conditions
 \begin{align}
		u_{\mathrm{B}}^\dagger u_{\mathrm{B}}=1_N\,,\quad u_{\mathrm{B}}^{\top}(\sigma_{2}\otimes1_{N/2})u_{\mathrm{B}}=\sigma_{2}\otimes1_{N/2}\,.
\end{align}
This means that the matrix $u_{\mathrm{B}}$ is an element
of the symplectic group ${\it Sp}(N/2)$. Therefore, the classifying
space of $U_{\mathrm{B}}$ is given by
  \begin{align}
    R_5= {\it Sp}(N/2)\,.
  \end{align}

\subsubsection{Class C}
The class C has the $\mathcal{C}_-$ symmetry with the condition
  \begin{align}
    C_{-}U_{\mathrm{B}}C_{-}^{-1}=-U_{\mathrm{B}},\,\quad C_{-}=(1_{N}\otimes i\sigma_{2})\mathcal{K}\,.
  \end{align}
Let  $\vec{F}^{(i)}\ (i=1,2,\cdots)$ are orthonormal eigenvectors with  $U_{\mathrm{B}}=+1$.
Then we can find that $C_{-}\vec{F}^{(i)}$  has the opposite eigenvalue $U_{\mathrm{B}}=-1$ from the above condition.
Therefore $U_{\mathrm{B}}$ has the same number of positive and negative eigenvalues, and the Witten index  in this class  is given by
  \begin{align}
    \Delta_{\mathrm{W}}=0\,.
  \end{align}

Let us consider the basis change such as 
	\begin{align}
			\widetilde C_{-}={\widetilde V}^{\top} C_{-}{\widetilde V}=(i\sigma_{2}\otimes 1_{N})\mathcal{K}\,,\ \quad {\widetilde V}=W_{N\leftrightarrow2}\,,
			\label{eq:basis_change_C}
	\end{align}
where $W_{N\leftrightarrow2}$ is defined by \eqref{eq:W_exchange}.
In this basis, we diagonalize the matrix $\widetilde U_{\mathrm{B}}\ (={\widetilde V}^{\top} U_{\mathrm{B}}{\widetilde V})$ as
 \begin{align}
	\widetilde{U}_{\mathrm{B}}=V
	\begin{pmatrix}
		1_{N} & 0\\
		0 & -1_{N}
	\end{pmatrix}
	V^{\dagger}\,,
	\label{eq:C-2}
\end{align}
where $V$ is a $2N\times2N$ unitary matrix which specifies the parameter space of ${U}_{\mathrm{B}}$. By using the redundancy of $V$,
the matrix can be given as follows :
 \begin{align}
	V=\left({\widetilde V}\vec{F}^{(1)},\cdots,{\widetilde V}\vec{F}^{(N)},-\widetilde C_{-}{\widetilde V}\vec{F}^{(1)},\cdots,-\widetilde C_{-}{\widetilde V}\vec{F}^{(N)}\right)\,.
	\label{eq:C-1}
\end{align}
From Eq.\:\eqref{eq:basis_change_C}, $V$ is of the form
  \begin{align}
    V=
      \begin{pmatrix}
        A & -B^{*}\\
        B & A^{*}
      \end{pmatrix}
    \label{eq:C-3}
  \end{align}
with $N\times N$ 
matrices $A$ and $B$.
Repeating the same argument in the class AII, the matrix $V$ defined in
this way is an element of the symplectic group ${\it Sp}(N)$.  
Furthermore, Eq.\:(\ref{eq:C-2})
is invariant under the transformation
  \begin{align}
    V\to V
      \begin{pmatrix}
        U & 0\\
        0 & U^{*}
      \end{pmatrix}
    \,,\quad U\in U(N)\,.
  \end{align}
Therefore, the classifying space of $U_{\mathrm{B}}$ is given by
  \begin{align}
    R_6=\frac{{\it Sp}(N)}{U(N)}\,.
  \end{align}

\subsubsection{Class CI}
The class CI has the three symmetries, $\mathcal{T}_+$, $\mathcal{C}_-$ and $\mathcal{T}_+\mathcal{C}_-$. The additional conditions for $U_{\mathrm{B}}$ are
  \begin{alignat}{2}
    T_{+}U_{\mathrm{B}}T_{+}^{-1}
    & =U_{\mathrm{B}},\qquad &T_{+}&=\mathcal{K}\,,
    \label{eq:T+_CI}
    \\
    C_{-}U_{\mathrm{B}}C_{-}^{-1}
    & =-U_{\mathrm{B}},\qquad &C_{-}&=(1_{N}\otimes i\sigma_{2})\mathcal{K}\,,\\
    \Gamma_{-}U_{\mathrm{B}}\Gamma_{-}^{-1}
    & =-U_{\mathrm{B}},\qquad&\Gamma_{-}&=1_{N}\otimes\sigma_{2}\,.
  \end{alignat}
Since the CS is present, the Witten index in the class CI
is
\begin{align}
	\Delta_{\mathrm{W}}=0\,.
\end{align}

When we take the basis
  \begin{align}
  	\widetilde\Gamma_{-}&={\widetilde V}^\dagger\Gamma_{-}{\widetilde V}=\sigma_{3}\otimes1_{N}\,,
  	\\
    \widetilde T_{+}&={\widetilde V}^\dagger T_+{\widetilde V}=(\sigma_{1}\otimes1_{N})\mathcal{K}=
      \begin{pmatrix}
        {0} & 1_{N}\cdot\mathcal{K}\\
        1_{N}\cdot\mathcal{K} & {0}
      \end{pmatrix}
  \end{align}
with the unitary matrix ${\widetilde V}$
\begin{align}
	&\widetilde V=W_{N\leftrightarrow 2}\frac{1}{\sqrt 2}
	\begin{pmatrix}
		1_N & 1_N\\
		i1_N & -i1_N
	\end{pmatrix}\,,
\end{align}
the matrix $\widetilde U_{\mathrm{B}}={\widetilde V}^\dagger U_{\mathrm{B}}{\widetilde V}$ is given by
  \begin{align}
    \widetilde{U}_{\mathrm{B}}=
      \begin{pmatrix}
        0 & u_{\mathrm{B}}\\
        u_{\mathrm{B}}^{\dagger} & 0
      \end{pmatrix}
  \quad u_{\mathrm{B}}^{\dagger}u_{\mathrm{B}}=1_{N}\,,
   \quad u_{\mathrm{B}}^{\top}=u_{\mathrm{B}}\,.
    \label{eq:CI-1}
  \end{align}
Since $u_{\mathrm{B}}$ is a unitary matrix, it can be diagonalized,
using the unitary matrix $v$, as
  \begin{align}
    u_{\mathrm{B}}=v^{\dagger}u_{\mathrm{d}}v\,,
    \quad u_{\mathrm{d}}=
      \begin{pmatrix}
        e^{i\alpha_{1}} &  & 0\\
        & \ddots\\
        0 &  & e^{i\alpha_{N}}
      \end{pmatrix}
    \,,
    \quad v\in U(N)\,,
    \quad\alpha_{i}\in\mathbb{R}\ \ \ (i=1,\cdots,N)\,.
    \label{eq:CI-2}
  \end{align}
From the relation $u_{\mathrm{B}}^{\top}=u_{\mathrm{B}}$,
we obtain
  \begin{align}
    (vv^{\top})u_{\mathrm{d}}=u_{\mathrm{d}}(vv^{\top})\,,
  \end{align}
or equivalently 
	\begin{align}
			(vv^{\top})_{ij}(u_{\mathrm{d}})_{jj}=(u_{\mathrm{d}})_{ii}(vv^{\top})_{ij}\,,
	\end{align}
where $i,j$ denote the indices of the matrix elements and are not summed  up.
If $(vv^{\top})_{ij}\neq0$, the above relation implies that$(u_{\mathrm{d}})_{ii}=(u_{\mathrm{d}})_{jj}$. 
Then, by dividing both sides of the above equation by $u_{\mathrm{d}}^{1/2}$, we can also obtain the following relation:
  \begin{align}
    (vv^{\top})_{ij}\sqrt{(u_{\mathrm{d}})_{jj}}=\sqrt{(u_{\mathrm{d}})_{ii}}(vv^{\top})_{ij}\,.
    \label{eq:CI-3}
  \end{align}
This relation is also trivially satisfied in the case of $(vv^{\top})_{ij}=0$.
Using this relation, $u_{\mathrm{B}}$ can be rewritten as 
  \begin{align}
    u_{\mathrm{B}}=w^{\top}w\,,
    \qquad
    w\equiv v^{\top}u_{\mathrm{d}}^{1/2}v^{*}\,,
    \label{eq:ub_CI}
  \end{align}
where $w$ satisfies
  \begin{align}
    ww^{\dagger}=1_N.
  \end{align}
This means the parameter space of $u_{\mathrm{B}}$ is specified by the unitary matrix $w$. However, Eq.\:(\ref{eq:ub_CI}) is invariant
under the replacement
  \begin{align}
    w\to gw\,,\quad g\in O(N)\,.
  \end{align}
Therefore, the classifying space of $U_{\mathrm{B}}$
is given by
  \begin{align}
    R_7=\frac{U(N)}{O(N)}\,.
  \end{align}

\subsection{$\mathbb{Z}_{2}$ index}
\label{subsec:Z2_index}
Finally, let us discuss the $\mathbb{Z}_{2}$ index, which is the number of Dirac zero modes in module 2.

\begin{figure}[h]
	\centering
		\includegraphics[height=4.5cm]{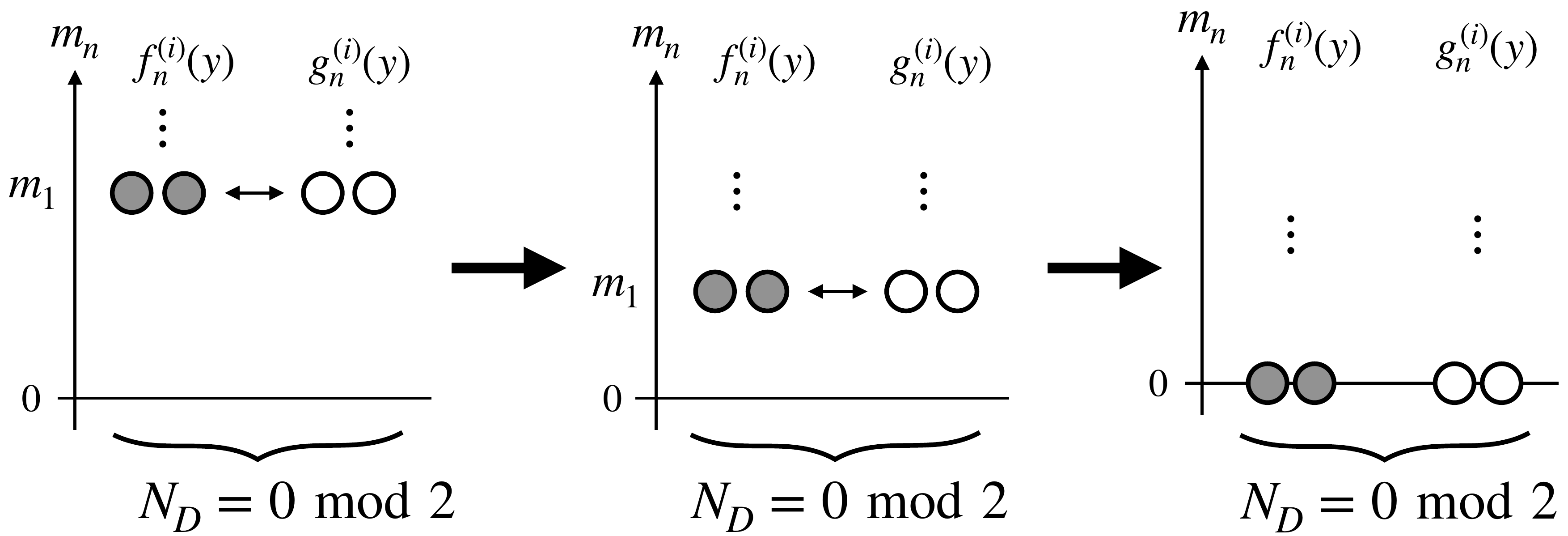}
		\ \\\vspace{1cm} 
		\includegraphics[height=4.5cm]{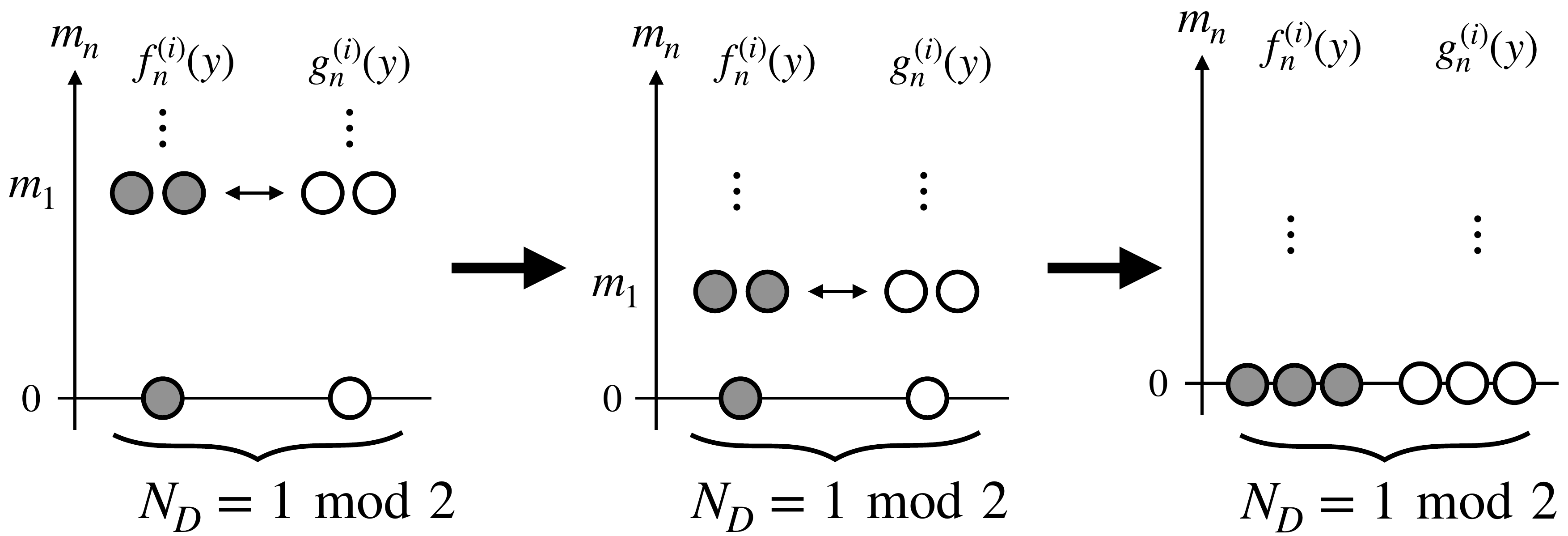}
	\captionsetup{width=.90\linewidth}
	\caption{The figures represent the change in the number of zero modes under continuous
		deformations of parameters with the symmetry $\mathcal{C}_+$. Comparing the top three figures with the
		bottom three figures, we can see that the number of massless Dirac fields $N_{D}$ mod 2 is invariant.}
	\label{figure:BDI-1}
\end{figure}

The  topological property of this index results from the degeneracy of massive mode functions due to the symmetry $\mathcal{C}_+$.
As we have seen in Section \ref{subsubsec:C+}, by the $\mathcal{C}_+$, the mode functions are transformed as follows:
  \begin{align}
   f_{n}^{(i)}(y)&\xrightarrow{\mathcal{C}_{+}}f_{n}^{\mathcal{C}_{+}(i)}(y)=Q_{y}P_{y}g_{n}^{(i)*}(y)\,,
   \\ 
   g_{n}^{(i)}(y)&\xrightarrow{\mathcal{C}_{+}}g_{n}^{\mathcal{C}_{+}(i)}(y)=Q_{y}P_{y}f_{n}^{(i)*}(y)\,.
\end{align}
Using Eqs.\:\eqref{eq:2-SUSYrelations1} and \eqref{eq:2-SUSYrelations2}, we can show that the $f_{n}^{(i)}(y)$ and $f_{n}^{\mathcal{C}_{+}(i)}(y)$ with the  same index $i$ are orthogonal with each other for $n\neq0$.
Furthermore, if we have a massive mode $f_{n}^{(j)}(y)\ (i\neq j)$ which is orthogonal with $f_{n}^{(i)}(y)$ and $f_{n}^{\mathcal{C}_{+}(i)}(y)$, $f_{n}^{\mathcal{C}_{+}(j)}(y)$ is also orthogonal with them.
This is also the same  for $g_{n}^{(i)}(y)$.
Therefore the degeneracy of the massive mode functions $f_{n}^{(i)}(y)$ and $g_{n}^{(i)}(y)\ (n\neq0)$ with the symmetry $\mathcal{C}_+$ is always multiple of two respectively, and their mass eigenvalues move together by deformations of the parameters.
On the other hand, zero mode functions $f_{0}^{(i)}(y)$ and $g_{0}^{(i)}(y)$ are not necessarily degenerate respectively (although the number of independent zero mode 
$f_{0}^{(i)}(y)$ and that of $g_{0}^{(i)}(y)$ are equal to each other by $\mathcal{C}_+$).  Then, the number of massless Dirac fields $N_D\ (\equiv (N_{f_0}+N_{g_0})/2)$ mod 2 is invariant by continuous deformations of the parameters in the boundary conditions as  described in Figure \ref{figure:BDI-1}, and this index can be topologically nontrivial if the symmetry $\mathcal{C}_+$ is present.

In the following subsection, we confirm that the $\mathbb{Z}_{2}$ index becomes topologically nontrivial and take the $\mathbb{Z}_{2}$ values in the class D and BDI depending on the discontinuity of their classifying spaces, while 
it is topologically trivial in the other classes.

\subsubsection{Class D}
First, let us consider the case of the class D. Since this class has the $\mathcal{C}_+$ symmetry, there is a possibility that the $\mathbb{Z}_{2}$ index becomes topologically nontrivial.
In this class, from the discussion in Section \ref{subsubsec:Witten index_D}, the matrix $U_{\mathrm{B}}$ can be written as follows:
	\begin{align}
			U_{\mathrm{B}}={\widetilde V} R(-\sigma_2\otimes1_N)R^{\top}{\widetilde V}^\dagger\,,\quad R\in O(2N)\,,
			\label{eq:UB_D}
	\end{align}
where
	\begin{align}
			{\widetilde V}=\frac{1}{\sqrt{2}}
			\begin{pmatrix}
				1_N & 1_{N/2}\otimes i\sigma_1
				\\
				-1_{N/2}\otimes i\sigma_2 & 1_{N/2}\otimes i\sigma_3
			\end{pmatrix}\,.
	\end{align}
The classifying space of this class is given by the disconnected space $O(2N)/U(N)$ from the matrix $R$ and divided into two connected regions by $\det R=+1$ and $\det R=-1$.
This determinant is related to the number of Dirac zero modes $N_D$.
The result is 
		\begin{align}
				\det R =(-1)^{N_D}\,.
				\label{eq:R_ND}
		\end{align}
Therefore, $N_D=0$ mod 2 for the parameter space with $\det R=+1$ and $N_D=1$ mod 2 for the parameter space with $\det R=-1$. This is the topological invariant as discussed in the beginning of this subsection, and we can find that this index corresponds to the topological number $\mathbb{Z}_2$ of the 0-th homotopy group of the classifying space.

To confirm this result, it is enough to see a simple example since this index is invariant by continuous deformations of parameters. Then, as an example, we take $R$ as
	\begin{align}
			R=
			\begin{pmatrix}
				1_{m}\otimes\sigma_3 & & &
				\\
				& 1_{N-2m} & &
				\\
				&&1_{2m}&
				\\
				&& &1_{N-2m}
			\end{pmatrix}	
			\ \ \ \ (m=0,\cdots,N/2)
	\end{align}
with $\det R=(-1)^m$.
For this $R$, $U_{\mathrm{B}}$ is given by
	\begin{align}
		U_{\rm B}&=
		 \left(\begin{array}{cc|cc}
			0_{2m}& & 1_{2m}&
			\\
			& 1_{N/2-m}\otimes\sigma_1 & & 0_{N-2m}
			\\\hline
			1_{2m}& & 0_{2m}&
			\\
			& 0_{N-2m}&&1_{N/2-m}\otimes\sigma_1
		\end{array}\right)
	\end{align}
Then the boundary condition \eqref{eq:BC-F} and \eqref{eq:BC-G}  for the zero modes are written as
	\begin{align} 
			&
		 \left(\begin{array}{cc|cc}
			1_{2m}& & -1_{2m}&
			\\
			& 1_{N/2-m}\otimes(1_2-\sigma_1) & & 0_{N-2m}
			\\\hline
			-1_{2m}& & 1_{2m}&
			\\
			& 0_{N-2m}&&1_{N/2-m}\otimes(1_2-\sigma_1)
		\end{array}\right)
		\begin{pmatrix}
			F_1^{(i)}e^{-M(L_{0}+\varepsilon)}
			\\
			F_1^{(i)}e^{-M(L_{1}-\varepsilon)}
			\\
			\vdots
			\\
			F_N^{(i)}e^{-M(L_{N-1}+\varepsilon)}
			\\
			F_N^{(i)}e^{-M(L_{N}-\varepsilon)}
		\end{pmatrix}
	=0\,,
	\\
		&  \left(\begin{array}{cc|cc}
			1_{2m}& & 1_{2m}&
			\\
			& 1_{N/2-m}\otimes(1_2+\sigma_1) & & 0_{N-2m}
			\\\hline
			1_{2m}& & 1_{2m}&
			\\
			& 0_{N-2m}&&1_{N/2-m}\otimes(1_2+\sigma_1)
		\end{array}\right)
	\begin{pmatrix}
		G_1^{(j)}e^{M(L_{0}+\varepsilon)}
		\\
		-G_1^{(j)}e^{M(L_{1}-\varepsilon)}
		\\
		\vdots
		\\
		G_N^{(j)}e^{M(L_{N-1}+\varepsilon)}
		\\
		-G_N^{(j)}e^{M(L_{N}-\varepsilon)}
	\end{pmatrix}
	=0\,,
	\end{align}
where the $F_a^{(i)}$ and $G_a^{(j)}\ (a=1,\cdots,N)$ are the coefficients in Eqs.\:\eqref{eq:zero mode f} and \eqref{eq:zero mode g}.
 Here we assume that the bulk mass $M\neq0$.
Then the coefficients are given by
	\begin{align}
	&F_{1}^{(i)}=F_{N/2+1}^{(i)}e^{M(L_{0}-L_{N/2})}\,,\ \ \cdots\,,\ \ F_{m}^{(i)}=F_{N/2+m}^{(i)}e^{M(L_{m-1}-L_{N/2+m-1})}\,,	
	\\
	&G_{1}^{(j)}=-G_{N/2+1}^{(j)}e^{-M(L_{0}-L_{N/2})}\,,\ \ \cdots\,,\ \ G_{m}^{(j)}=-G_{N/2+m}^{(j)}e^{-M(L_{m-1}-L_{N/2+m-1})}\,,	
\end{align}
and the others vanish for the case of $M\neq 0$. Therefore, the independent coefficients are $F_{1}^{(i)},\cdots,F_{m}^{(i)}$ and $G_{1}^{(i)},\cdots,G_{m}^{(i)}$, and
the number of independent zero mode functions is  $m$ for $f_0^{(i)}\ (i=1,\cdots,m)$ and  $g_0^{(j)}\ (j=1,\cdots,m)$, respectively.
Thus $N_D=m$ for the case of $M\neq0$ and we can find that the relation \eqref{eq:R_ND} is satisfied. 
We also mention that we obtain $N_D=N-m$ Dirac zero modes for $M=0$ unlike the case of $M\neq0$. But the result \eqref{eq:R_ND} does not change since $N$ is even in this class.

\subsubsection{Class BDI}
Next, we consider the class BDI. This class has the $\mathcal{T}_+$ symmetry in addition to $\mathcal{C}_+$. From the discussion in Section \ref{subsubsec:Witten index_BDI},
$U_{\mathrm{B}}$ can be written as follows:
	\begin{align}
		U_{\mathrm{B}}={\widetilde V}\begin{pmatrix}0 & u_{\mathrm{B}}\\
			u_{\mathrm{B}}^{\top} & 0
		\end{pmatrix}{\widetilde V}^{\top}\,,\quad u_{\mathrm{B}}\in O(N)\,,
	\label{eq:UB_BDI}
	\end{align}
where
	\begin{align}
	{\widetilde V}=\frac{1}{\sqrt2}
	\begin{pmatrix}
		1_N & 1_{N/2}\otimes\sigma_1
		\\
		-1_{N/2}\otimes i\sigma_2 & 1_{N/2}\otimes\sigma_3
	\end{pmatrix}\,.
\end{align}
We mention that if we restrict the matrix $R$ in \eqref{eq:UB_D} as 
\begin{align}
	R=\begin{pmatrix}
		u_{\mathrm{B}} & 0
		\\
		0 & 1_N
	\end{pmatrix}\,,
\end{align}
Eq.\:\eqref{eq:UB_BDI} can be obtained.

The classifying space in this class is $O(N)$ specified by the matrix $u_{\mathrm{B}}$ and  divided into two spaces by $\det u_{\mathrm{B}}=+1$ and $\det u_{\mathrm{B}}=-1$ disconnected with each other.
By considering the same example in the class D, we obtain the following relation between this determinant and the number of Dirac zero modes $N_D$:
	\begin{align}
			\det u_{\mathrm{B}}=(-1)^{N_D}\,.
	\end{align}
Therefore, we see that the $\mathbb{Z}_2$ index in this class corresponds to the topological number $\mathbb{Z}_2$ of the 0-th homotopy group of this classifying space.

\subsubsection{The other classes}
In the class DIII, the $\mathcal{T}_-$ symmetry is present in addition to $\mathcal{C}_+$.
Under the $\mathcal{T}_-$ transformation, the mode functions are transformed as 
\begin{align}
	f_{n}^{(i)}(y)&\xrightarrow{\mathcal{T}_{-}}f_{n}^{\mathcal{T}_{-}(i)}(y)=Q_{y}f_{n}^{(i)*}(y)\,,
	\\ 
	g_{n}^{(i)}(y)&\xrightarrow{\mathcal{T}_{-}}g_{n}^{\mathcal{T}_{-}(i)}(y)=Q_{y}g_{n}^{(i)*}(y)\,.
\end{align}
Then we can show that the $f_{n}^{(i)}(y)\ (g_{n}^{(i)}(y))$ and $f_{n}^{\mathcal{T}_{-}(i)}(y)\ (g_{n}^{\mathcal{T}_{-}(i)}(y))$ with the same index $i$ are orthogonal to each other for all $n$.
Therefore this $\mathcal{T}_-$ symmetry leads to the twofold degeneracy of  not only the massive modes  but also zero modes like the Kramers degeneracy.
Then the number of zero mode $N_D$ mod 2 is always trivial in this class.
Instead of the $\mathbb{Z}_2$ index,
one may consider  a fourfold degeneracy of massive mode functions by the combination of $\mathcal{T}_-$ and $\mathcal{C}_+$ and $N_D$ mod 4 to become topologically nontrivial,  but this is not the case since the fourfold degeneracy of massive modes does not generally hold. 
Actually, when we consider the boundary condition 
	\begin{align}
			U_{\mathrm{B}}&=
		 \left(\begin{array}{ccc|ccc}
				u_1 &  &&&&
			\\ &\ddots&&&0&
			\\
			&&u_{N/2}&&&
			\\\hline
			&&&u_1 &  &
			\\ &0&&&\ddots&
			\\
			&&&&&u_{N/2}
		\end{array}\right)\,,
	\\\\
	u_a&=\begin{pmatrix}
		\cos\theta_a & \sin\theta_a
		\\
		\sin\theta_a & -\cos\theta_a
		\end{pmatrix}\,,
	\qquad\theta_a\in[0,2\pi)\qquad
	(a=1,\cdots,N/2)\,,
	\end{align}
two Dirac zero modes  appear for each $\theta_a$ which satisfies $\tan(\theta_a/2)=e^{-M(L_a-L_{a-1})}$.
Then we find that the Dirac zero modes vanish by continuous deformations of the parameters $\theta_a\ (a=1,\cdots,N/2)$, and thus it is  topologically trivial.

We can also show that the $\mathbb{Z}_2$ index is trivial in the other classes since they have no symmetries which leads to the degeneracy only for massive modes.

\section{Summary and Discussion}
\label{sec:Summary and Discussion}

In this paper, we studied 5d fermions of which extra dimension is on quantum graphs.
We showed that the boundary conditions of the quantum graphs are classified into ten symmetry classes according to the presence or absence of time-reversal, charge conjugation, and extra-spatial symmetries of 5d fermions. The obtained symmetry classes are identical to the AZ symmetry classes of SPT phases of gapped free-fermion systems.
A Hermitian matrix $U_{\mathrm{B}}$ specifying the boundary conditions corresponds to a 0d Hamiltonian of gapped free fermion systems. 
Furthermore, the constraints for $U_{\mathrm{B}}$ originating from symmetries of 5d fermions are 
the same as those for the 0d Hamiltonian with AZ symmetries.
Based on these results, we introduced  topological numbers for the boundary conditions in the same manner as those of 0d topological insulators and superconductors. 
Importantly, the topological numbers of the boundary conditions coincide with the number of  KK 4d chiral or Dirac fermions localized at the vertex of quantum graphs, which 
would be a generalization of the bulk-boundary correspondence for gapped free-fermion systems. 

Our classification implies that  class A with no symmetries or class AI  with the time-reversal symmetry is preferable for realizing the fermion flavor structure in the standard model.
The nontrivial topological number $\mathbb{Z}$ in these classes may provide three generations of  4d chiral fermions.\footnote{
For example, Refs.~\cite{Fujimoto:2012wv,Fujimoto:2013ki,Fujimoto:2014fka,Fujimoto:2017lln,Fujimoto:2019lbo}  discuss the realization of the mass hierarchies and CP-violating phase by using  the 1d extra space consisting of three line segments with the Dirichlet boundary conditions for fermions. This corresponds to the case with $|\Delta_W|=3$, and thus three generations of 4d chiral fermions appear. }
To fully reproduce the standard model in our quantum graph approach, we must investigate gauge fields on quantum graphs.
(Higgs bosons can be obtained by five-dimensional gauge fields.)
Although Refs.~\cite{Fujimoto:2012wv,Fujimoto:2013ki,Fujimoto:2014fka,Fujimoto:2017lln,Fujimoto:2019lbo} take into account gauge fields, they consider the only simple boundary conditions such as the Dirichlet and Neumann conditions.
In general graphs,  more involved boundary conditions are possible for gauge fields. 
The boundary conditions for gauge fields  are related to those of fermions by 5d gauge symmetry since the 5d fermions after the gauge transformation should satisfy the same boundary condition, which restricts an allowed gauge parameter space. 
This restriction affects the 4d spectra of the gauge fields and can induce gauge symmetry breaking.
We should also take care of the gauge anomalies due to the 4d chiral fermions,
but our quantum graph approach is based on a 5d theory, so it should be anomaly-free. 
The anomalies of 4d chiral fermions at the vertex of quantum graphs are canceled with the contributions of massive modes at the edges of quantum graphs by  the anomaly inflow mechanism~\cite{Callan:1984sa,Witten:2015aba,Witten:2019bou}.

The key point in the classification of the boundary conditions for the 5d fermions
is the existence of chiral spinors in four dimensions.
The hermiticity of the boundary matrix $U_{\mathrm{B}}$ originates from the independence of the left and right-handed chiral fields under the 4d Lorentz invariance.
We expect that the same discussion works for general odd $D$-dimensional cases since 
chiral spinors exist in $D-1$ dimensions.
In contrast, for even $D$-dimensions, we cannot apply the same discussions. In fact, the matrix for the boundary conditions is not generally Hermitian in even dimensions (See e.g. \cite{Inoue:2021euj}).
We leave the extension of our results to other spacetime dimensions as a future problem. In particular, the investigations for lower dimensions would be important from the viewpoint of condensed matter physics.

We are also interested in the correspondence to SPT phases in other dimensions.
So far, we have discussed the relation between the  boundary conditions for quantum graphs and  1+0d SPT phases for gapped free fermion systems, assuming that the 4d spacetime and the extra dimension are factorized, and the boundary conditions on the extra space do not have a 4d momentum dependence.
However, if we consider the boundary conditions depending on the 4d momentum (or equivalently the 4d derivative), the boundary condition matrix would be regarded as a higher dimensional Hamiltonian, and thus we may obtain the correspondence to SPT phases in other dimensions.

Another future direction is the effect of interactions.
So far, we have considered free fermions on quantum graphs. However, by considering interactions and also quantum corrections, the boundary conditions would be changed, and breakdowns of the topological phases may occur, which affects low-energy physics. 
For the topological matter side, it has been known that SPT phases of gapped  free-fermion systems  may break down by interactions preserving symmetries (see e.g.
\cite{Fidkowski:2009dba,PhysRevB.83.075103,Ryu:2012he,Fidkowski:2013jua,Wang:2014lca,You:2014sqa,Morimoto:2015lua,Kapustin:2014dxa}).
For example, the $\mathbb{Z}$ classification of class BDI for 1d topological superconductor  reduces to the $\mathbb{Z}_8$ classification if quartic interactions are present.
We are interested in whether a similar breakdown occurs in the topological classification of boundary conditions.
It is also known that interacting SPT phases in bulk can be characterized by perturbative and/or nonperturbative anomalies on boundary \cite{Ryu:2010ah,Wen:2013oza,Hsieh:2015xaa,Witten:2015aba,Witten:2016cio}. This is  described by an anomaly inflow. 
Thus, it could be interesting to consider the boundary conditions from the viewpoint of anomalies. 
We hope to revisit these issues in the near future.

\section*{Acknowledgements}
This work is supported by Japan Society for the Promotion of Science
(JSPS) KAKENHI Grant Nos. JP21J10331, JP20H00131 and JP18K03649 and JST CREST Grant No. JPMJCR19T2, and JST ERATO-FS Grant No. JPMJER2105.

\small
\setlength{\itemsep}{-2pt}
\bibliographystyle{utphys}
\bibliography{ref-1}

\end{document}